%% file: Boltzmann_bedload.tex
\documentclass[aps,prl,twocolumn,superscriptaddress]{revtex4-1}

\usepackage{graphicx}

\input{./exemplary_experiment.tex}

\begin{document}

\title{Boltzmann distribution of sediment transport}

\author{A.~Abramian}
\author{O.~Devauchelle}
\email[]{devauchelle@ipgp.fr}
\affiliation{Institut de Physique du Globe de Paris, 1 rue Jussieu, 75238 Paris, France}
\author{G.~Seizilles}
\affiliation{Institut de Physique du Globe de Paris, 1 rue Jussieu, 75238 Paris, France}
\affiliation{Now at Naxicap Partners, 5-7 rue de Monttessuy, 75007 Paris, France}
\author{E.~Lajeunesse}
\affiliation{Institut de Physique du Globe de Paris, 1 rue Jussieu, 75238 Paris, France}

\date{\today}

\begin{abstract}
The coupling of sediment transport with the flow that drives it allows rivers to shape their own bed. Cross-stream fluxes of sediment play a crucial, yet poorly understood, role in this process. Here, we track particles in a laboratory flume to relate their statistical behavior to the self organization of the granular bed they make up. As they travel downstream, the transported grains wander randomly across the bed's surface, thus inducing cross-stream diffusion. The balance of diffusion and gravity results in a peculiar Boltzmann distribution, in which the bed's roughness plays the role of thermal fluctuations, while its surface forms the potential well that confines the sediment flux.
\end{abstract}

\pacs{
47.57.Gc, % Granular flow
05.40.Jc, % Brownian motion
92.40.Pb 	% Geomorphology
}

\maketitle

% \section{Introduction}

When water flows over a layer of solid grains, the shear stress it exerts on the sediment's surface entrains some of the grains as bedload \cite{bagnold1973nature,mouilleron2009inside}. Eventually, the flow deposits the traveling grains downstream \cite{charru2004erosion, lajeunesse2010bed}. The balance of entrainment and deposition deforms the sediment bed \cite{exner1925uber}, thus changing the flow and the distribution of shear stress. This coupling, through various instabilities, generates sand ripples in streams \cite{coleman2000sand,charru2006ripple}, rhomboid patterns on beaches \cite{devauchelle2010rhomboid}, alternate bars in rivers \cite{colombini1987finite} and, possibly, meanders \cite{ikeda1981bend,johannesson1989linear, liverpool1995dynamics,seminara2006meanders}.

More fundamentally, the coupling of water flow and sediment transport is the mechanism by which alluvial rivers choose their own shape and size, as they build their bed out of the sediment they carry \cite{henderson1961stability,parker2007physical,devauchelle2011longitudinal,seizilles2013width}. To do so, however, rivers need to transport sediment not only downstream, but also across the flow \cite{parker1978self,parker1978self2}. On a slanted bed, of course, gravity will pull traveling grains downwards; it thus diverts the sediment flux away from the banks of a river \cite{kovacs1994new,chen2009sediment}. What mechanism opposes this flux to maintain the river's bed remains an open question. Here, we suggest the inherent randomness of sediment transport plays a major role in the answer.

The velocity of bedload grains fluctuates as they travel over the rough bed, and the bedload layer constantly exchanges particles with the latter \cite{roseberry2012probabilistic}, thus calling for a statistical description of bedload transport \cite{jenkins1998collisional,furbish2012probabilistic}. At its simplest, this theory involves a population of non-interacting grains traveling, on average, at velocity $V_x$, close to the grain's settling velocity \cite{lajeunesse2010bed,lajeunesse2017bedload}. If $n$ is the surface density of traveling grains, the downstream flux of sediment reads $q_s = n V_x$. As long as sediment transport is weak, the traveling grains do not interact significantly, and their average velocity $V_x$ can be treated as a constant. Both the streamwise and cross-stream velocities, nonetheless, fluctuate significantly \cite{roseberry2012probabilistic,seizilles2014cross}.

\begin{figure}[h!]
\includegraphics[width = 1\columnwidth]{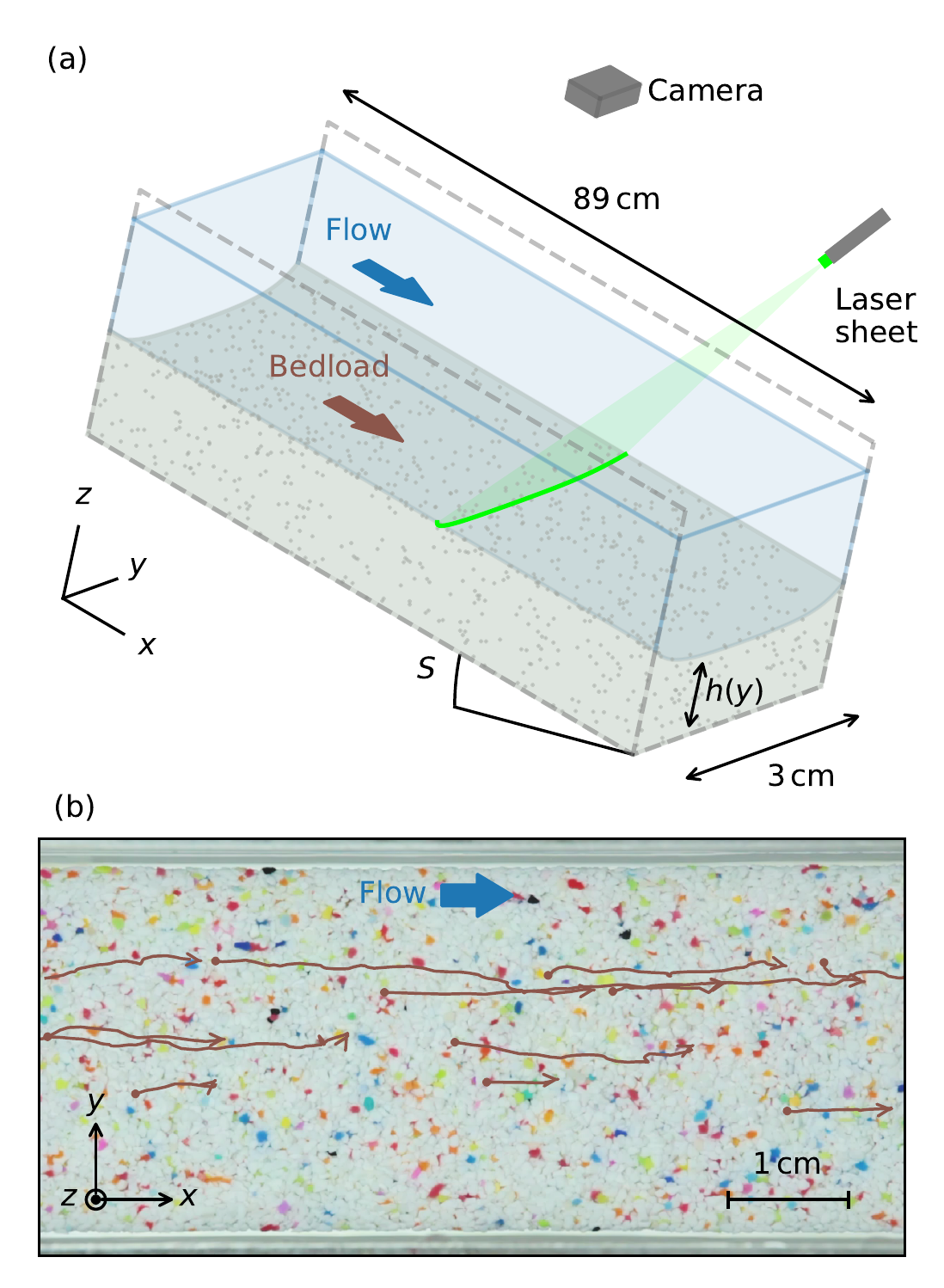}
\caption{\label{fig:setup}(a)~Experimental setup and notations. Two Plexiglas panels confine the flume laterally. The $x$ axis is aligned with the flow. (b)~Part of the camera's field of view (background picture) with superimposed grains trajectories (red lines). Dots and arrows indicate beginning and end of trajectories, respectively. Data from experimental run~\run.}
\end{figure}
\begin{table}
\caption{\label{tab:exp_para}Experimental parameters. Run {\run} serves as an example in all figures.}
\begin{ruledtabular}

\begin{tabular}{ccccc}
Run & Sediment input & Fluid input & Slope & Tracking time \\
\# & [grains$\,$s$^{-1}$] & [L$\,$min$^{-1}$] & \% & [min] \\
\hline
\underline{1} & 42.4 & 0.83 & 0.88 & 93 \\
2 & 37.4 & 1.12 & 0.79 & 100 \\
3 & 21.3 & 0.87 & 0.77 & 181 \\
4 & 19.7 & 1.13 & 0.69 & 68 \\
5 & 19.2 & 1.11 & 0.71 & 124 \\
\end{tabular}

\end{ruledtabular}
\end{table}
A little-investigated consequence of these fluctuations is the cross-stream dispersion they induce \cite{nikora2002bed,aussillous2016scale}. Indeed, as it travels downstream, a grain bumps into immobile grains like a ball rolling down a Galton board. The random deviations so induced turn its trajectory into a random walk across the stream \cite{samson1998diffusive,seizilles2014cross}. We thus expect a cross-stream, Fickian flux to bring traveling grains towards the less populated areas of the bed (lower $n$). Mathematically,
\begin{equation}
  q_d = - \ell_d \frac{\partial q_s}{\partial y}
  \label{eq:diffusion_flux}
\end{equation}
where $q_d$ is the fluctuation-induced Fickian flux, $y$ is the cross-stream coordinate, and $\ell_d$ is the diffusion length, which scales with the amplitude of the trajectory fluctuations. Tracking resin grains in a water flume, Seizilles \emph{et al.} found $\ell_d \approx 0.03\,d_s$ ($d_s$ is the grain size) \cite{seizilles2014cross}. To our knowledge, neither the cross-stream flux of grains $q_d$, nor its consequences on the bed's shape, have been directly observed.

% \section{Laboratory flume}

To measure the Fickian flux $q_d$, we set up a particle-tracking experiment in a 3$\,$cm-wide flume \{Fig.~\ref{fig:setup}(a), detailed experimental methods in Sup.~Mat.\}. We inject into the flume a mixture of water and glycerol (density $\rho= 1160\,$g$\,$L$^{-1}$, viscosity $\eta= 10\,$cP) at constant rate (Tab.~\ref{tab:exp_para}). We use a viscous fluid to keep the flow laminar (Reynolds number below 250). Simultaneously, and also at constant rate, we inject sieved resin grains (median diameter $d_s= 827\,\mu$m, density $\rho_s=1540\,$g$\,$L$^{-1}$). After a few hours, the sediment bed reaches its equilibrium shape.

This equilibrium, however, is a dynamical one: the flow constantly entrains new grains, and deposits other ones onto the bed. A camera mounted above the flume films the traveling grains through the fluid surface, at a frequency of 50$\,$fps [Tab.~\ref{tab:exp_para}, Fig.~\ref{fig:setup}(b), Sup.~Movie]. Although made of the same material, the grains are of different colors, which allows us to locate them individually on each frame. We then connect their locations on successive frames to reconstruct their trajectories with a precision of $0.1\,d_s$ \cite[Sup.~Mat.]{munkres1957algorithms}.

\begin{figure}[h!]
\includegraphics[width = 1\columnwidth]{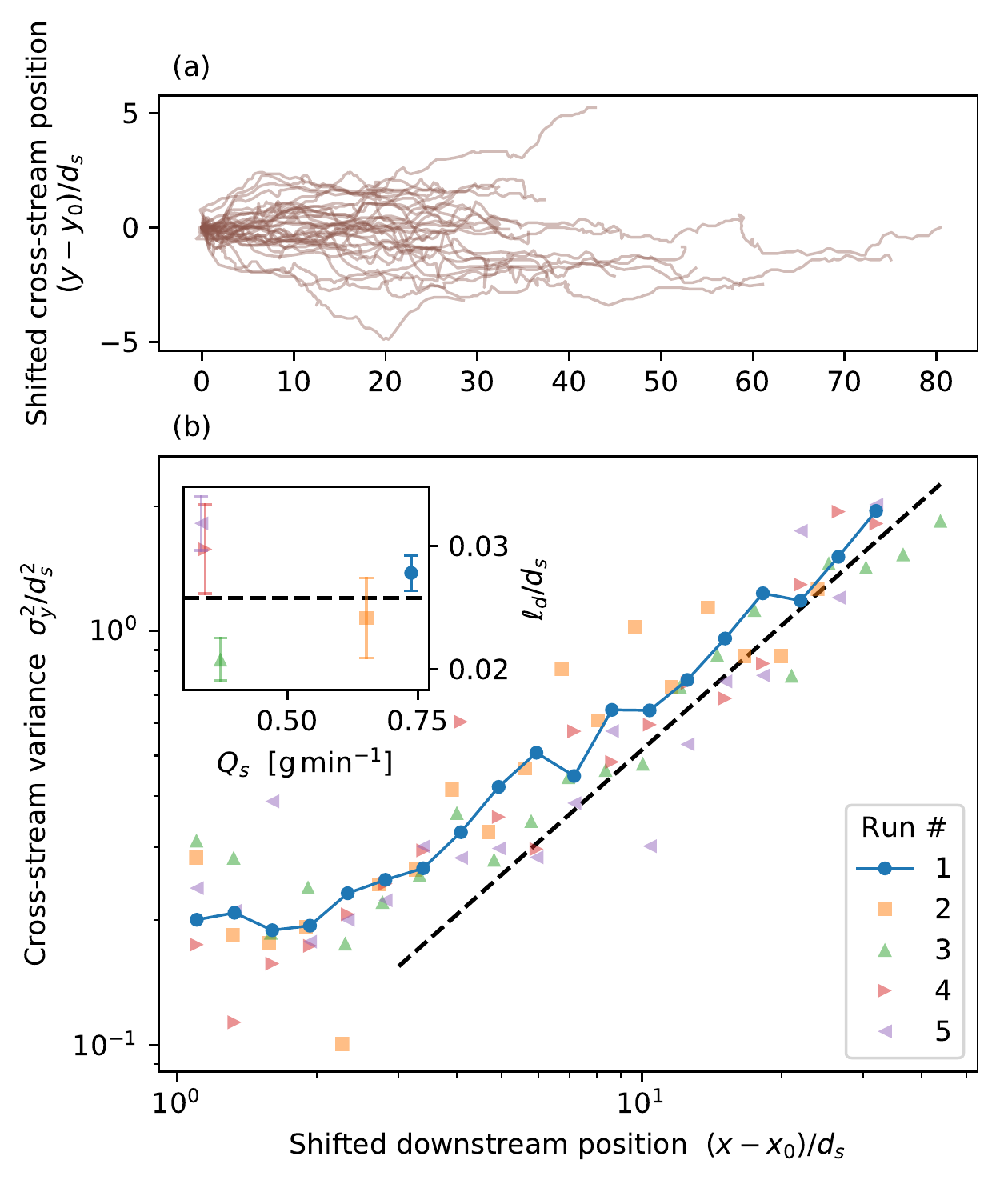}
\caption{Cross-stream dispersion of the traveling grains. (a) 34 trajectories from run~\run, with starting point shifted to origin. (b) Cross-stream variance of shifted trajectory. Dashed black line: linear relation with diffusion length $\ell_d$ fitted to data [Eq.~(\ref{eq:diff_length})]. Inset: $\ell_d$ fitted independently to individual runs. Error bars show uncertainty. \label{fig:beloadDiffusion}}
\end{figure}

The resulting trajectories are mostly oriented downstream, as expected, but they also fluctuate sideways, like in previous bedload experiments \citep{nikora2002bed,roseberry2012probabilistic,seizilles2014cross}. These fluctuations cause them to disperse across the stream as they travel downstream [Fig.~\ref{fig:beloadDiffusion}(a)]. We now distribute all our trajectories into 25 logarithmically-spaced bins, according to their travel length $x-x_0$ ($x_0$ is the starting point of each trajectory), and calculate the variance $\sigma_y^2$ for each bin [Fig.~\ref{fig:beloadDiffusion}(b)]. We find that, for trajectories longer than a few grain diameters, the cross-stream variance increases linearly with the travel distance. \citet{seizilles2014cross} interpreted a similar relationship as the signature of a random walk across the stream (they also found that the auto-correlation of the cross-stream trajectories decays exponentially). Accordingly, we now fit the relation $\sigma_y^2 = 2 \ell_d ( x - x_0 )$ to our trajectories (beyond $3 d_s$ downstream of their starting point). To estimate $\ell_d$, we treat the above relation as the reduced major axis of our data set: $2 \ell_d = \mathrm{std}(\sigma_y^2)/\mathrm{std}( x - x_0 )$, where the standard deviation is over bins, that is, over the data points of Fig.~\ref{fig:beloadDiffusion}(b). Using our entire data set (typically $3\times10^4$ trajectories per run), we get
\begin{equation}
  \ell_d = \left( 0.024 \pm 0.002 \right) d_s \, .
  \label{eq:diff_length}
\end{equation}
where the uncertainty is the expected standard deviation of $\ell_d$. This value is close to previous measurements in pure water \cite{seizilles2014cross}, although it is most likely affected by the physical properties of the fluid and of the grains.

That diffusion expresses itself through a length scale, as opposed to a diffusion coefficient, betrays its athermal origin: it is the driving (here, the flow) that sets the time scale. This property relates bedload diffusion to the diffusion induced by shearing in granular materials and foams \cite{utter2004self,fenistein2004universal,debregeas2001deformation}.

Although fluctuations disperse the traveling grains across the stream, most grains travel near the center of the channel [Fig.~\ref{fig:setup}(b)]. According to Eq.~\ref{eq:diffusion_flux}, such a concentration should induce a Fickian flux towards the flume's wall. At equilibrium, we expect gravity to counteract this flux; for this to happen, the bed's cross section needs to be convex. In fact, the bed cannot maintain a flat surface: in this configuration, the fluid-induced shear stress would be weaker near the sidewalls, and so would  the intensity of sediment transport. Bedload diffusion would then bring more grains towards the bank, thus preventing equilibrium. In the following, we investigate this coupling.

\begin{figure}
\includegraphics[width = .99\columnwidth]{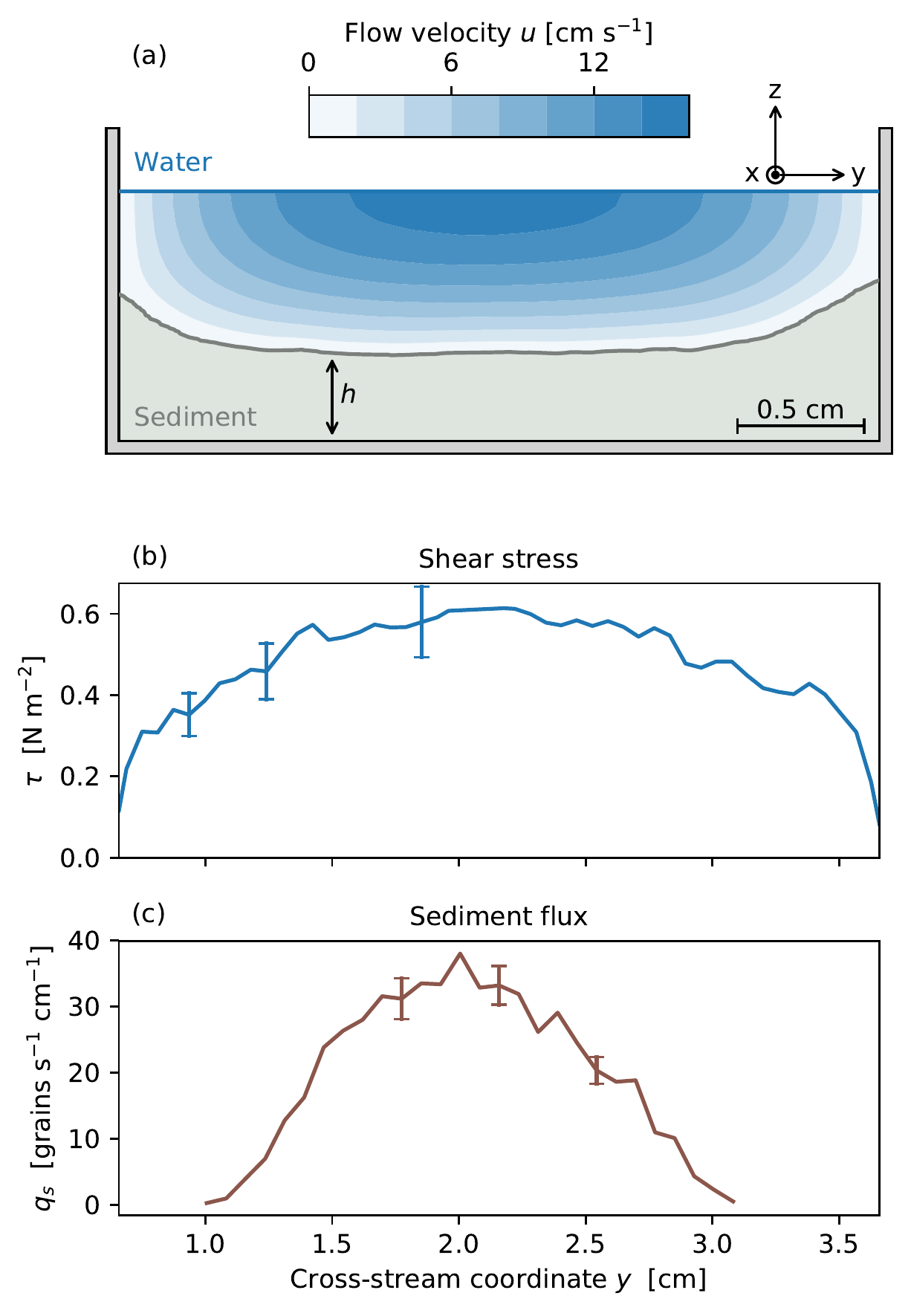}
\caption{\label{fig:cross_section}(a)~Average cross-section of the flume during run~\run. Beige: sediment layer. Blue colormap: downstream flow velocity calculated with finite elements. (b)~Flow-induced shear stress on the bed. (c) Sediment flux measured by grain tracking. Error bars indicate measurement uncertainty (Sup.~Mat.).}
\end{figure}

We use an inclined laser sheet to measure the elevation profile of the bed [Fig.~\ref{fig:setup}(a), Sup.~Mat]. The laser source is fixed on a rail which allows it to scan the flume over 20$\,$cm. We evaluate the tilt of the rail by scanning a tub of still milk; it is less than $0.03\,$\%. At the end of an experimental run, we switch off the fluid input; this brings the bed to a standstill in a matter of seconds. We then let the fluid drain out of the flume, and use the laser scanner (i) to measure the bed's downstream slope $S$ (Tab.~\ref{tab:exp_para}) and (ii) to spatially average the bed's cross-section, $h(y)$ [Fig.~\ref{fig:cross_section}(a)]. We find that the sediment bed is convex for all our experimental runs. Its surface gently curves upwards near the center of the flume, and steepens near the walls. This observation indicates that the sediment bed has spontaneously created a potential well to confine the traveling particles in its center.

% \section{Flow}

Unfortunately, measuring the flow depth based on the deflection of the laser beam proved imprecise. Instead, we used finite elements to solve the Stokes equation in two dimensions \cite[Sup.~Mat.]{MR3043640}, namely:
\begin{equation}
  \eta \nabla^2 u = \rho g S \, ,
\end{equation}
where $u$ is the streamwise velocity of the fluid. We further assume that the free surface is flat, that the viscous stress vanishes there ($\partial u/\partial z = 0$, where $z$ is the vertical coordinate) and that the fluid does not slip at the bed's surface ($u=0$). Knowing the bed's downstream slope $S$, we then adjust the elevation of the water surface to match the fluid discharge. In addition to the flow depth, this computation provides us with the velocity field of the flow [Fig.~\ref{fig:cross_section}(a)], and thus the intensity of the viscous stress $\tau$ that the fluid exerts on the bed [Fig.~\ref{fig:cross_section}(b)]. We find that, like the sediment bed, the viscous stress varies across the flume; it reaches a maximum at the center of the channel, and vanishes where the bed's surface joins the walls---as expected for a laminar flow.

We now wish to relate the flow-induced stress to sediment transport. To measure the latter, we divide the flume's width into 50 bins, and count the trajectories that cross a constant-$x$ line within each bin, per unit time. This procedure yields a sediment-flux profile (Sup.~Mat.). Repeating it for 10 different lines across the channel, we obtain an average sediment-flux profile, $q_s(y)$ [we keep only data points for which the relative uncertainty is less than one, Fig.~\ref{fig:cross_section}(c)]. In accordance with the distribution of trajectories in Fig.~\ref{fig:setup}(b), the sediment flux appears concentrated around the center of the flume. It vanishes quickly away from the center, much before the fluid-induced stress has significantly decreased.

Following Shields, we now relate the sediment flux to the ratio of the fluid-induced stress to the weight of a grain, $\theta$ \cite{shields1936anwendung}:
\begin{equation}
  \theta = \frac{ \tau }{ ( \rho_s - \rho ) g d_s }
  \label{eq:Shields}
\end{equation}
where $g$ is the acceleration of gravity. The Shields parameter is an instance of the Coulomb friction factor; strictly speaking, on a convex bed like that of Fig.~\ref{fig:cross_section}(a), its expression should include the cross-stream slope, $\partial h/\partial y$ \cite{seizilles2013width}. In our experiments, however, we find that this correction is insignificant where sediment transport is measurable. Accordingly, we content ourselves with the approximate expression of Eq.~(\ref{eq:Shields}).

\begin{figure}
\includegraphics[width = .99\columnwidth]{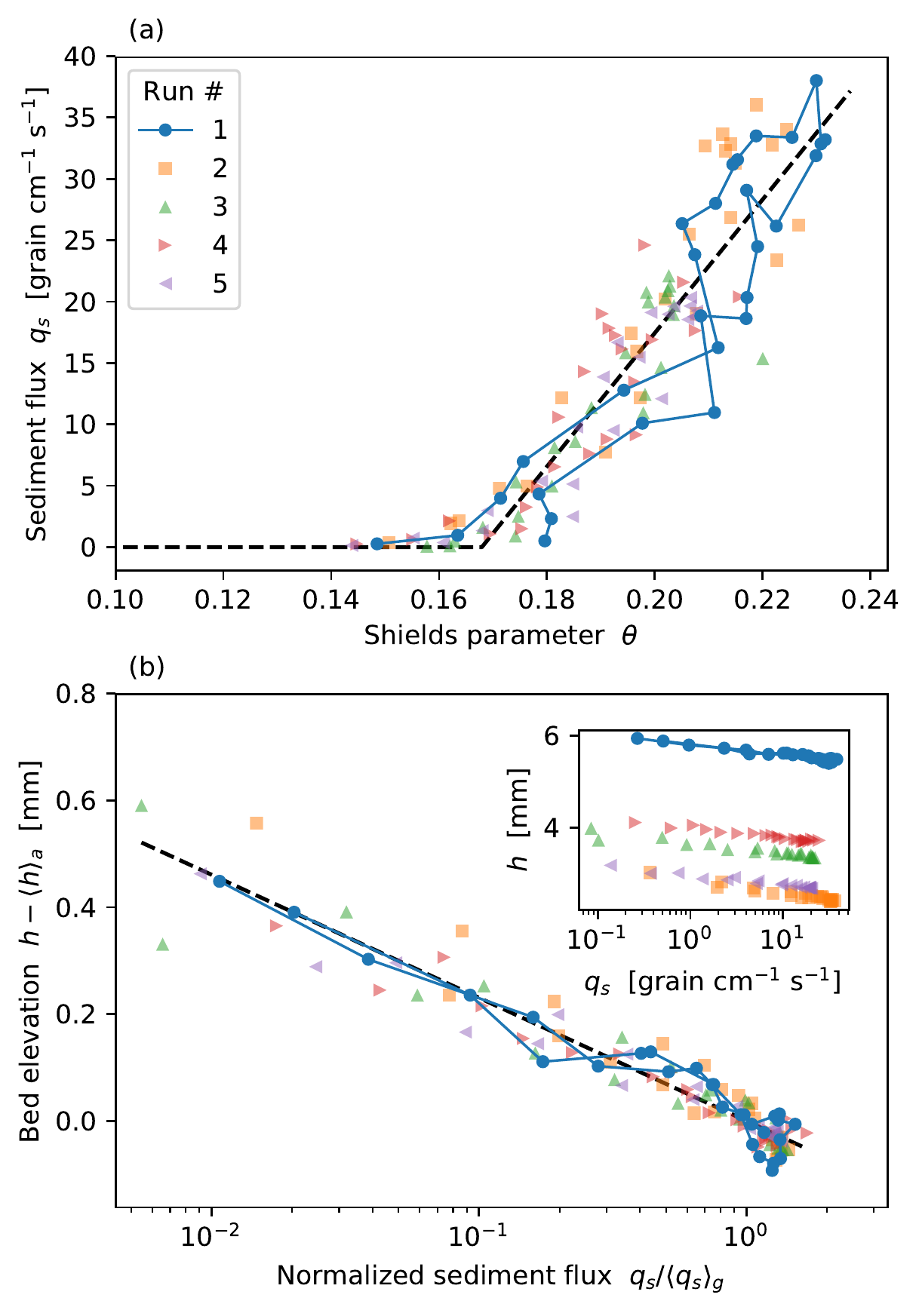}
\caption{\label{fig:Boltzmann}(a)~Local sediment transport law. Marker types indicate individual experimental runs. Solid line: experimental run~\run. Dashed black line: Equation~(\ref{eq:transport_law}) with $q_0 = 544$~grains~s$^{-1}$~m$^{-1}$ and $\theta_t=0.17$. (b)~Distribution of sediment flux with respect to bed elevation. Colors and markers similar to (a). Dashed black line: Bolztmann distribution [Equation~(\ref{eq:Boltzmann})] with $\lambda_B=0.10$ mm.}
\end{figure}

Plotting the intensity of the sediment flux as a function of the force driving it, in the form of the Shields parameter $\theta$, shows a well-defined threshold [Fig.~\ref{fig:Boltzmann}(a)]: no grain moves when the fluid-induced stress is too weak to overcome its weight, but the sediment flux increases steeply past this threshold. This emblematic behavior, apparent in a single experimental run, is confirmed by the superimposition of our five experimental runs [Fig.~\ref{fig:Boltzmann}(a)]. Indeed, within the variability of the measurements, the five corresponding transport laws gather around a common relation, which we may treat as linear above the threshold Shields stress $\theta_t$ \cite{seizilles2014cross}:
\begin{equation}
  q_s = q_0 \, ( \theta - \theta_t )
  \label{eq:transport_law}
\end{equation}
where $q_0$ is a constant of order $ ( \rho - \rho_s ) g / \eta $. Fitting this transport law to our complete data set, we get
$ q_0 = 544 \pm 48$~grains~s$^{-1}$~cm$^{-1}$ and $\theta_t = 0.167 \pm 0.003 $, where the uncertainty is the standard deviation over individual runs. These values correspond to a typical transport law in a laminar flow \cite{mouilleron2009inside,seizilles2014cross}.

The local intensity of the flow-induced stress controls the local flux of sediment---just as expected. More surprisingly, perhaps, the sediment bed needs to adjust its shape so that, in total, the flume conveys the sediment discharge that we impose at the inlet. We suggest that it does so by balancing the Fickian flux, $q_d$, which pushes the traveling grains away from the flume's center, with the gravity-induced flux, $q_g$, which pulls them towards the lowest point of the bed's surface. As a first approximation, we may assume that the latter is proportional (i) to the cross-stream slope of the bed and (ii) to the local intensity of the downstream flux of sediment, $q_s$. Mathematically,
\begin{equation}
  q_g = - \alpha \, q_s \frac{\partial h}{\partial y}
  \label{eq:gravity_flux}
\end{equation}
where $\alpha$ is a dimensionless constant. Although conducted in air, the experiments of \citet{chen2009sediment} suggest that it should be of order unity or less.

At equilibrium, the gravity-induced flux $q_g$ needs to match the Fickian flux $q_d$. Adding Eqs.~(\ref{eq:gravity_flux}) and (\ref{eq:diffusion_flux}) yields the Boltzmann equation, which we readily integrate into an exponential distribution:
\begin{equation}
  q_s (y) = q_0 \, \exp \left[ -\frac{ h(y) }{\lambda_B} \right] \, ,
  \label{eq:Boltzmann_dim}
\end{equation}
where $q_0$ is an integration constant, and $\lambda_B=\ell_d/\alpha$ is the characteristic length of the distribution. Distinctively, this distribution relates two quantities ($q_s$ and $h$) that depend on the space coordinate $y$, but the latter does not explicitly appear in its expression. This, however, does not make it a local relationship: unlike the transport law of Eq.~(\ref{eq:transport_law}), it features an integration constant which depends on the sediment and water discharges of each experiment. These properties, typical of a Boltzmann distribution, appear when plotting the bed elevation as a function of the sediment discharge [Fig.~\ref{fig:Boltzmann}(b), inset]. For each experiment, the data points trace twice the same line in the semi-logarithmic space, as they go from one side of the channel to the other, but the position of this line depends on the experimental run.

To bring all our experiments into the same space, we now divide Eq.~(\ref{eq:Boltzmann_dim}) by its geometrical mean. This rids us of the integration constant $q_0$, and turns the distribution of sediment transport into
\begin{equation}
  \frac{q_s(y)}{\langle q_s \rangle_g } = \exp \left[ -\frac{h(y) - \langle h \rangle_a}{\lambda_B} \right] \, ,
  \label{eq:Boltzmann}
\end{equation}
where $\langle \cdot \rangle_g$ and $\langle \cdot \rangle_a$ are the geometric and arithmetic means, respectively. Within the variability of our observations, the data points from all experimental runs gather around a straight line, which we interpret as Eq.~(\ref{eq:Boltzmann}). Fitting the characteristic length $\lambda_B$ to our entire data set, we find $\lambda_B=0.10 \pm 0.01\,$mm, where the uncertainty is the standard deviation over individual runs. More tellingly, this value corresponds to
\begin{equation}
  \lambda_B = (0.12 \pm 0.02) \, d_s \, ,
\end{equation}
showing that the characteristic length compares with the grain size. Returning to the definition of $\lambda_B$, we find that the constant $\alpha$ in Eq.~(\ref{eq:gravity_flux}) is about $0.2$, in agreement with previous estimates \cite{chen2009sediment}.

Although temperature plays no part here, the structure of  Eq.~(\ref{eq:Boltzmann}), as well as its derivation, makes it a direct analog of the Boltzmann distribution, where the cross-stream deviations of the grains' trajectories play the role of thermal fluctuations. Pursuing this analogy, we suggest that the scale of $\lambda_B$ is inherited from the roughness of the underlying granular bed, which we believe causes the cross-stream deviations---a mechanism reminiscent of, but somewhat simpler than, the shear-induced diffusion observed in granular flows \cite{utter2004self,fenistein2004universal,debregeas2001deformation}. To support this hypothesis, however, we would need more experiments with different grains and fluids.

% \section{Conclusion and discussion}

The familiarity of the Boltzmann distribution should not obscure the peculiarity of the phenomenon we report here. We naturally expect that random walkers will distribute themselves in a potential well according to this distribution; what is remarkable here, however, is that the system spontaneously chooses the shape of the potential well to match the transport law. This is possible only because the sediment bed is made of the very particles that roam over its surface.

A practical consequence of this self-organization is that sediment transport cannot be uniform across a flume, thus prompting us to reevaluate the transport laws measured in this classical set-up. (If we were to assume uniformity in our experiments, we would underestimate $q_0$ by a factor of two.) In the context of dry granular flows, the traditional rotating-drum experiment has been challenged on similar grounds \cite{jop2005crucial}.

Much remains to be done to understand how the bed builds its own shape. To do so, we will have to drop the equilibrium assumption. A first step in that direction was to demonstrate theoretically that the cross-stream diffusion of sediment could generate a distinctive instability, but the associated pattern has not been observed yet \cite{abramian2018}. More generally, the consequences of bedload diffusion on the morphology of rivers, and ultimately on that of the landscapes they carve, belong to uncharted territory.

\begin{acknowledgments}
We thank F.~M\'etivier, D.H.~Rothman and A.P.~Petroff for insightful discussions. We are also indebted to P.~Delorme for the sediment feeder, and to J.~Heyman for his suggestions on particle tracking. O.D. and E.L. were partially funded by the \emph{\'Emergence(s)} program of the City of Paris, and the EC2CO program, respectively.
\end{acknowledgments}

\bibliography{Boltzmann_bedload_bibliography}

\end{document}

\title{Boltzmann distribution of sediment transport\\
  \emph{Supplemental Material}
  }

\author{A.~Abramian}
\author{O.~Devauchelle}
\email[]{devauchelle@ipgp.fr}
\affiliation{Institut de Physique du Globe de Paris, 1 rue Jussieu, 75238 Paris, France}
\author{G.~Seizilles}
\affiliation{Institut de Physique du Globe de Paris, 1 rue Jussieu, 75238 Paris, France}
\affiliation{Now at Naxicap Partners, 5-7 rue de Monttessuy, 75007 Paris, France}
\author{E.~Lajeunesse}
\affiliation{Institut de Physique du Globe de Paris, 1 rue Jussieu, 75238 Paris, France}

\date{\today}

\begin{abstract}
We report here the experimental methods used to produce the results presented in the main document. More details can be found in reference \cite{theseAnais}\footnote{\url{hal.archives-ouvertes.fr/tel-01992434v1}}, Ch.~2.
\end{abstract}

\maketitle

\section{Grains and fluid}

\begin{figure}
\centering
\includegraphics[width = \columnwidth]{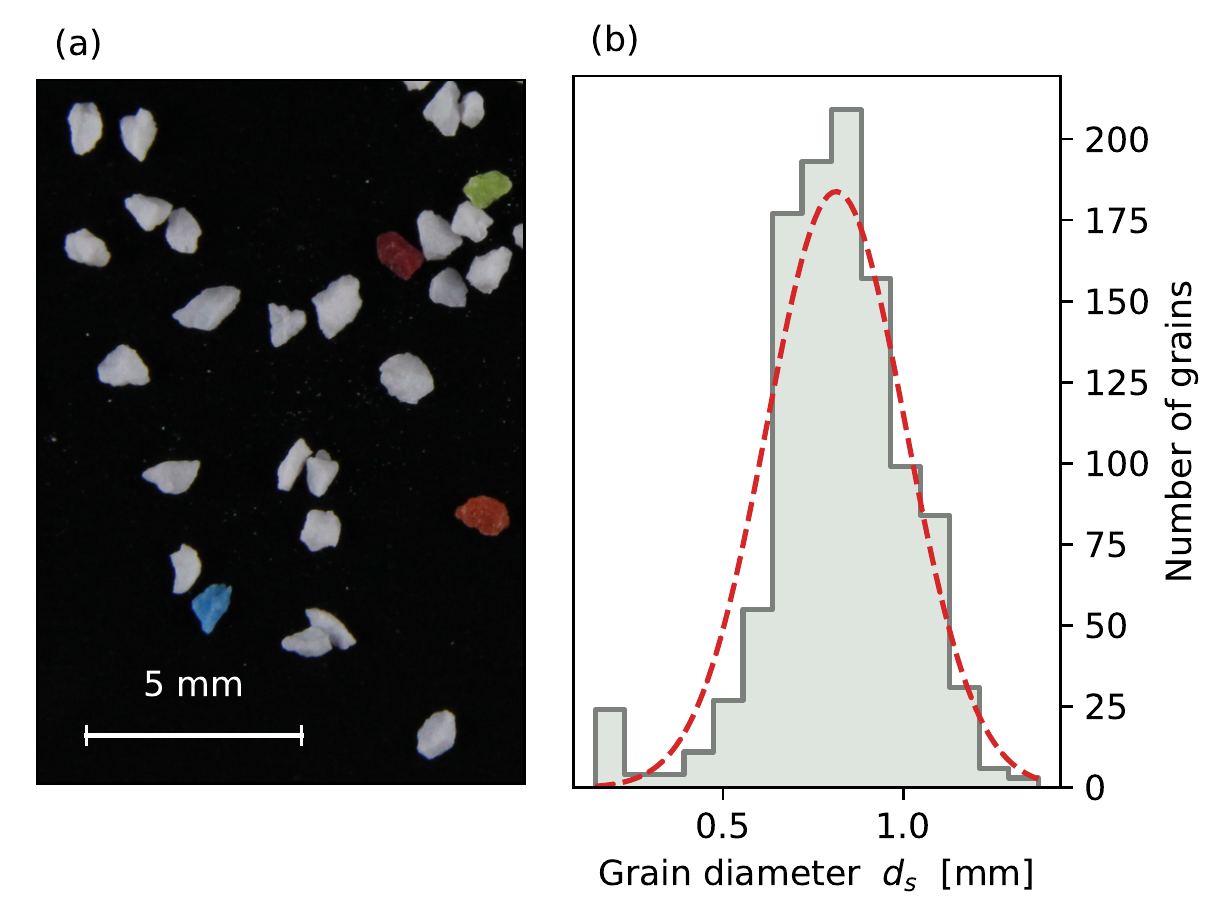}
\caption{(a) Sediment grains. (b) Shaded histogram: grain-size distribution of a sample containing 1084 grains. Red dashed line: normal distribution with the mean and standard deviation of the sample.\label{fig:grain_size_distribution}}
\end{figure}

In this article, all experiments involve the same irregular resin grains, a blast media manufactured by Guyson\footnote{\url{www.guyson.co.uk}} [Fig.~\ref{fig:grain_size_distribution}(a)]. \citet{seizilles2013width} used the same material to produce laboratory rivers. The grains are made of urea resin, the density of which is 1520$\,$kg$\,$m$^{-3}$ (manufacturer's value). Using a pycnometer, we find an average density of $\rho_s=$1488$\,$kg$\,$m$^{-3}$; we use this value in our numerical calculations.

The manufacturer provides a grain-size range of $[650,800]\,{\mathrm \mu}$m. Using the software ImageJ\footnote{\url{imagej.nih.gov/ij/}}, we analyze a picture similar to Fig.~\ref{fig:grain_size_distribution}(a) to measure the grain-size distribution of Fig.~\ref{fig:grain_size_distribution}(b). Representing each particle as a sphere, we find an average grain diameter of $0.81\pm0.19\,$mm (standard deviation), and a median diameter of $d_s=0.82\,$mm (the value we use in the paper).

The advantage of resin over quartz or glass is its low density, which reduces the settling velocity of the grains, and therefore slows them down as they travel in the bedload layer \citep{seizilles2014cross}. This slowness facilitates the grain-tracking procedure described in Sec.~\ref{sec:grain_tracking}. We estimate this settling velocity with Stokes' formula for the settling velocity of a sphere:
\begin{equation}
  V_s = \dfrac{( \rho_s - \rho ) g d_s^2}{18 \rho \nu } \,,
\end{equation}
the value of which depends on the fluid's density $\rho$ and viscosity $\nu$.

To further slow down our grains (and to prevent turbulence), we use a mixture of water (60$\,$\%) and glycerol (40$\,$\%) to entrain the grains. As water can evaporate over the course of an experimental run, we measure the density of our mixture with an oscillating U-tube densimeter (Anton Paar\footnote{\url{www.anton-paar.com}} DMA 5000 M) at the beginning of each run, and before each measurement. We compensate for evaporation by adding water to the fluid to maintain its density near $\rho = 1161\,$g$\,$L$^{-1}$ with an accuracy of about $1.3\,$g$\,$L$^{-1}$ (standard deviation over experiments). This ensures that its viscosity remains near $\rho \nu = ( 1.02 \pm 0.03 )\times10^{-2} \,$Pa$\,$s$^{-1}$.

Based on the above values, we find that the settling velocity of our grains is about $V_s= 1.0\,$cm$\,$s$^{-1}$, with a relative uncertainty of about $50\,$\%, mainly due to the dispersion of the grain size. We thus expect the traveling grains to move at about 12 times their own length per second; this frequency is a lower bound for the frame rate of the movies used for grain tracking.

\section{Grain tracking}\label{sec:grain_tracking}

\begin{figure}
\centering
\includegraphics[width = \columnwidth]{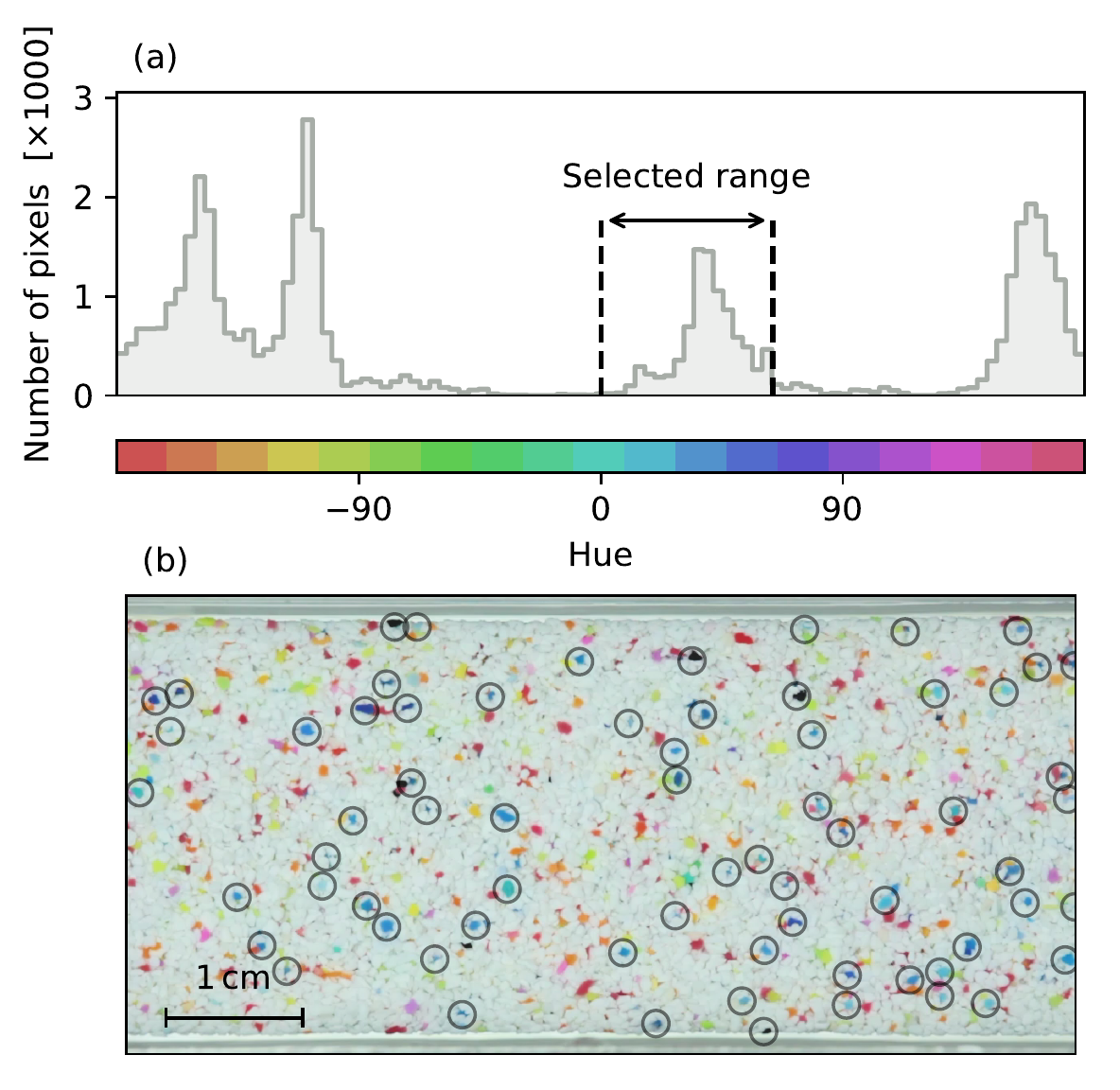}
\caption{Grain detection. (a) Histogram of pixel hue for the frame shown in (b). Selected range (black arrow) corresponds to blue grains. (b) Movie frame from run~{\run} (Tab.~I, main document). Dark circles: detected blue grains. Actual picture is broader than shown.
}
\label{fig:histogram_hues}
\end{figure}

Most of our grains are white, but some of them come in a distinct hue: mainly red, orange or blue (a few percent of the total for each color) [Fig.~\ref{fig:grain_size_distribution}(a)]. We use this property to track them using top-view movies of the traveling grains (Fig.~\ref{fig:histogram_hues}), recorded with a Canon\footnote{\url{global.canon/en/}} 700D SLR camera, fitted with a 250$\,$mm macro lens. This camera allows us to reccord movies at 50 color frames per second. Each frame is a $1280\times1024$ array; at this resolution, the grain size $d_s$ is about 10 pixels.

To locate individual grains on a frame, we first decompose the image array into its hue, saturation and value. We then select pixels whose saturation is more than 0.3, to ensure that their hue is well-defined. We then choose a hue range (for instance $[0^{\circ},64^{\circ}]$ for blue grains) and pick the pixels that fall into this range [Fig.\ref{fig:histogram_hues}(a)]. When submitted to these criteria, a movie frame yields an array of booleans, where clusters corresponds to blue grains. We smooth this array by convolving it with a 10-pixel Gaussian filter, and locate the maximums of the resulting frame using the \verb|peak_local_max| function of the Scikit-image\footnote{\url{scikit-image.org}} Python library, requiring a minimum distance of 10 pixels between peaks.

Figure~\ref{fig:histogram_hues}(b) shows the result of this grain-detection procedure for the movie frame shown in Fig.~1 (main document). In total, we find 74 blue grains in this frame. Assuming each of them occupies an area $d_s^2$, we estimate that blue grains make up about $1.8\,$\% of the sediment. This value, of course, depends on the definition of ``blue'' (the hue range).

At this point, we can only speculate that the uncertainty on the position of each grain resulting from the detection procedure lies between a pixel (about $0.1 d_s$) and $d_s$. To estimate it more accurately, we first need to track the grains through a series of movie frames, thus reconstructing their trajectories.

To track a particle over two successive frames, we use a variant of the Hungarian algorithm \cite{munkres1957algorithms,ancey2014microstructural}, implemented in the Munkres\footnote{\url{pypi.org/project/munkres/}} Python library. We often loose track of a traveling particle as it drifts near a bed particle of the same color; we thus allow a particle to disappear during 10 successive frames before assigning its position to a new trajectory.

Figure 1(b) (main document) shows some trajectories measured with the above procedure, but the movie provided as supplemental material gives a better sense of the result. Before counting traveling grains to measure a sediment flux (Sec.~\ref{sec:sed_flux}), we first turn to immobile grains to estimate the uncertainty of our tracking procedure. We select, among all the trajectories of run~\run, those which lasted at least 1 minute (500 frames), and extend downstream over less than 10 pixels (about $d_s$). There are 101 such trajectories, which we assume correspond to immobile grains belonging to the sediment bed. On average, their standard deviation is $0.41\,$px along the stream, and $0.52\,$px across the stream; we therefore estimate that the uncertainty on the position of each grain is about one pixel.

\section{Sediment flux}\label{sec:sed_flux}

\begin{figure}
\centering
\includegraphics[width = \columnwidth]{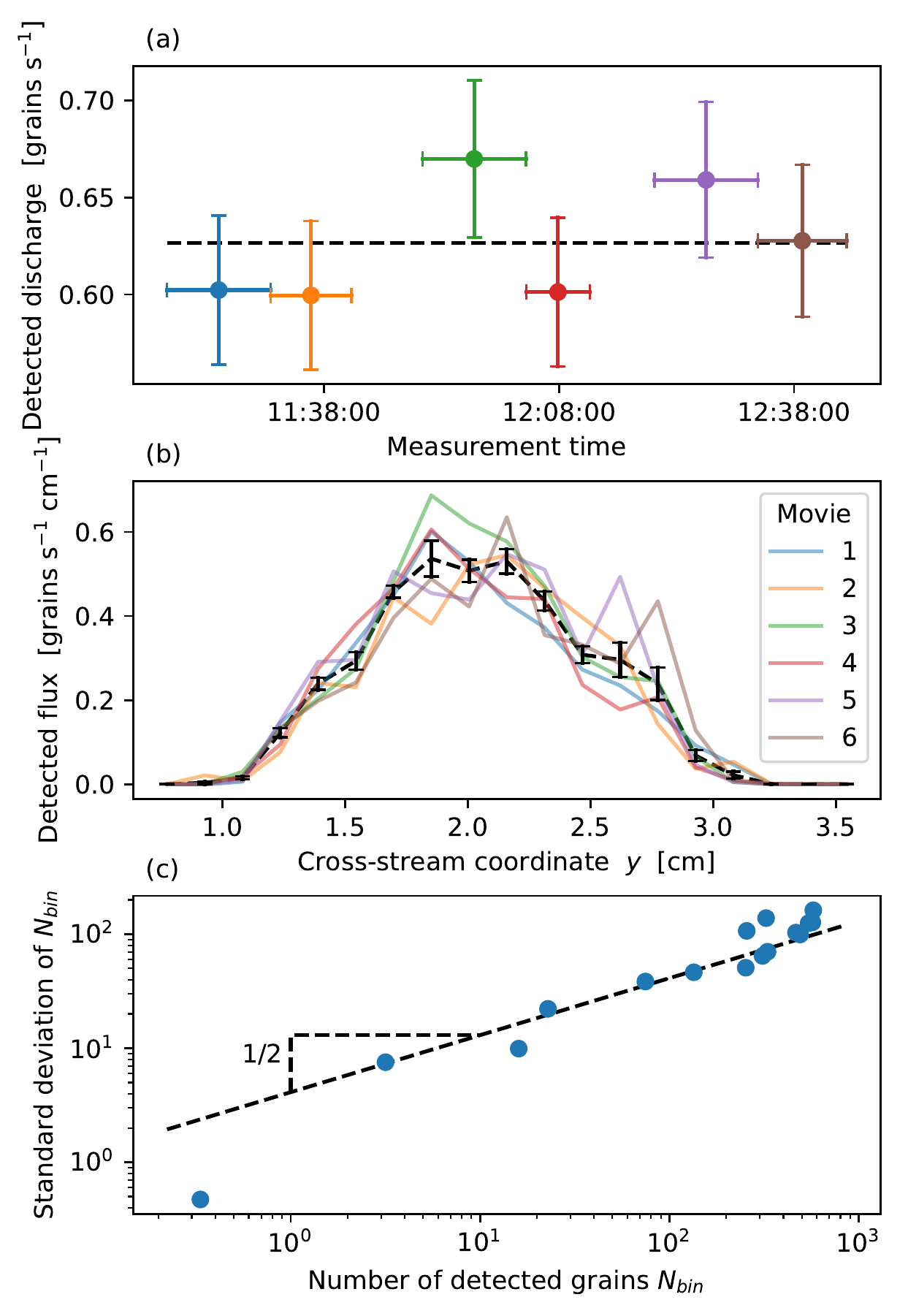}
\caption{Sediment flux measurement for run~\run. (a) Total discharge of blue grains, for 6 independent movies. Horizontal error bars correspond to movie duration. The experiment was started the day before time on horizontal axis. Vertical error bars: uncertainty based on Eq.~\eqref{eq:std_N_cross}. Dashed black line: average over movies. (b) Flux of blue particles. Colors correspond to (a). Dashed black line: average over movies. Black error bars: uncertainty based on Eq.~\eqref{eq:std_N_cross}. (c) Standard deviation of the number of detected crossings, as a function of the number of detected crossings (blue dots). Dashed black line: fitted square-root relation.\label{fig:steady_sediment_transport}}
\end{figure}

To translate a collection of trajectories into a sediment flux, we first draw $N_{lines}=10$ evenly spaced lines across the stream, and find where they intersect the trajectories. We record the cross-stream coordinate $y$ of each intersection, as well as its direction [upstream ($+1$) or downstream ($-1$)]. A typical movie records between 4,000 and 12,000 intersections during its $\Delta t \approx 10\,$min duration. Summing the directions of these intersections, and dividing the result by the number of cross-stream lines, and by the movie duration, yields the discharge corresponding to a collection of trajectories [Fig.~\ref{fig:steady_sediment_transport}(a)]. For the experimental run~\run, we have recorded 6 movies in about an hour, and tracked the blue particles in each of them. This provides us with six independent measurement of the discharge for the same experimental run. Although these measurements fluctuate around their average, they do not show any clear trend, indicating that the system is in steady state. The amplitude of the fluctuations (about 10$\,$\%) matches the expected deviation for a normally distributed number of intersections [vertical error bars in Fig.~\ref{fig:steady_sediment_transport}(a)].

Averaged over the 6 movies of run~\run, the discharge of blue grains is about $0.63\,$grains$\,$s$^{-1}$, while the sediment input for this run is $42.4\,$grains$\,$s$^{-1}$. The ratio of the two yields an estimate of the proportion of blue grains: about $1.5\,$\%, to be compared with the value of $1.8\,$\% found in Sec.~\ref{sec:grain_tracking}. This mismatch might be due to the difference between the numerical proportion of a class of grains, and the proportion of visible area it occupies on the bed's surface. To bypass this issue, we normalize the sediment flux profiles of Fig.~3 (main document) with the total sediment input.

We now use the location of the intersections between trajectories and cross-stream lines to calculate the cross-stream profile of the sediment flux. Before summing the directions of the intersections, we first distribute them into 19 even-sized bins according to their position across the stream (only 16 were populated during run~\run). The discharge through a bin, divided by the bin size, yields an estimate of the local sediment flux for each movie [Fig.~\ref{fig:steady_sediment_transport}(b)]. Like with the total discharge, we find that the sediment flux profile fluctuates around the average, but show no significant trend.

Finally, we now have six independent measurements of the sediment flux trough $N_{bin} = 19$ bins across the stream; we may thus probe the statistics of the sediment flux measurement [Fig.~\ref{fig:steady_sediment_transport}(c)]. For each bin, we count the average number of recorded crossings, $N_{cross}$, and its standard deviation (both quantities are with respect to the 6 independent measurements). The statistical significance of such a small number of independent measurements is certainly questionable, but we nonetheless find that the classical square root relationship between standard deviation and sample size tolerably fits our data:
\begin{equation}
  \sigma[N_{cross}] \approx 4.1 \,\sqrt{ \langle N_{cross} \rangle  } \, ,
  \label{eq:std_N_cross}
\end{equation}
where the standard deviation, $\sigma$, and the mean, $\langle \cdot \rangle$, are with respect to the 6 independent measurements. We further assume that $\langle N_{cross} \rangle$ is proportional to the local sediment flux:
\begin{equation}
  \langle N_{cross} \rangle \approx q_s \, \Delta t \, N_{lines} \, \dfrac{W}{N_{bin}} \,
  \label{eq:N_cross_qs}
\end{equation}
where $W$ is the width of the channel. Equations~\eqref{eq:std_N_cross} and \eqref{eq:N_cross_qs} yield an approximate formula for the uncertainty on the sediment flux [error bars on Figs.~\ref{fig:steady_sediment_transport}(a) and \ref{fig:steady_sediment_transport}(b), and on Fig.~3 (main document)].

\section{Cross section}

\begin{figure}
\centering
\includegraphics[width = \columnwidth]{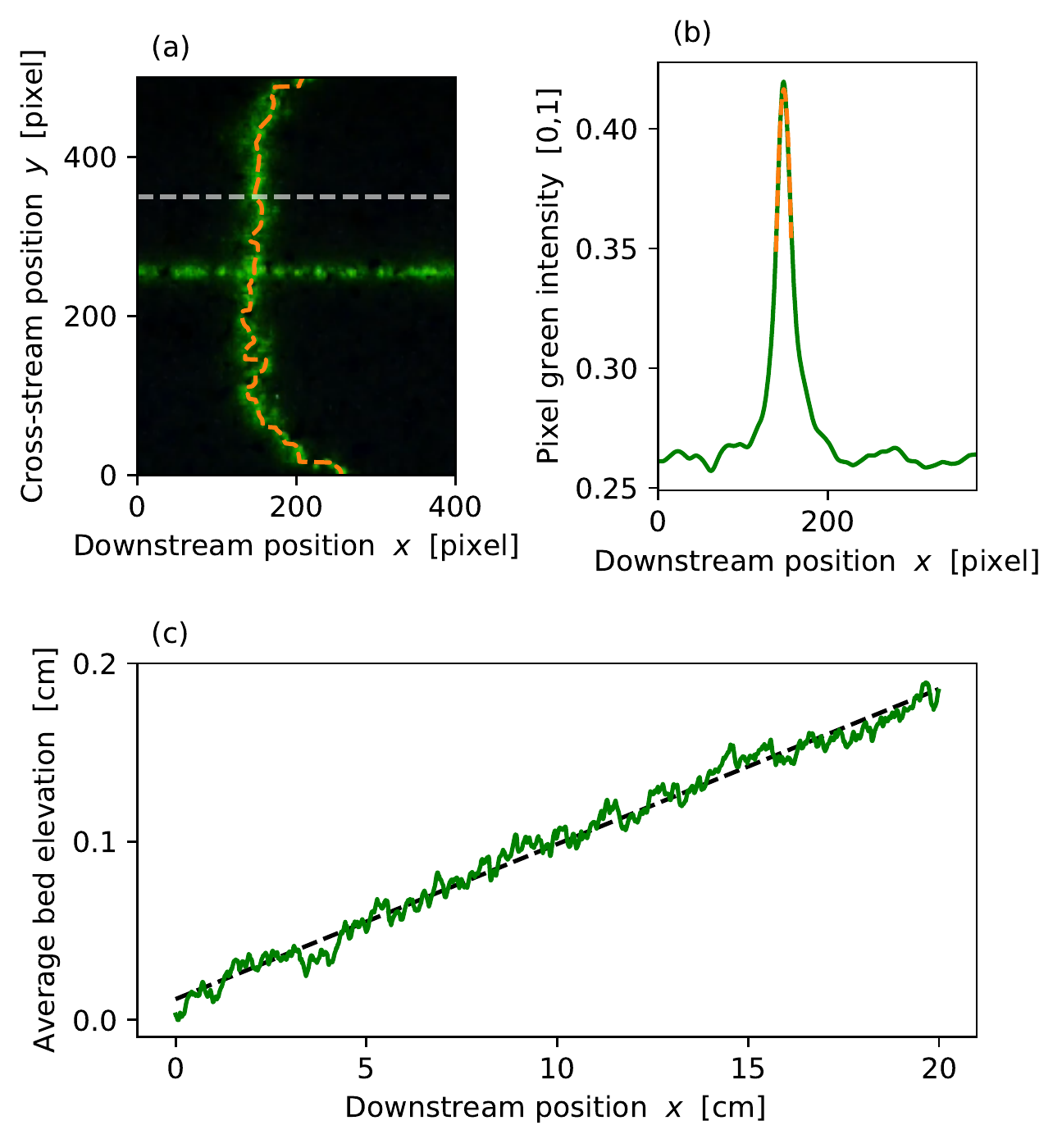}
\caption{Bed-elevation measurement. (a) Laser sheet projected on the channel bed (top view), after smoothing with a 4-pixel Gaussian filter. Orange dashed line: location of peak intensity. (b) Green intensity of pixels along the white, dashed line of (a) ($y = 570\,$pixels). Orange dashed line: fit of the laser peak with a second-order polynomial. (c) Cross-stream average of the sediment bed elevation measured as a function of the position of the laser carriage (green line). Dashed black line: affine fit.\label{fig:laser_line}}
\end{figure}

At the end of an experimental run, we switch off the water and sediment inputs. The moving grains settle down and stop within seconds of the run's end. We then let the fluid drain out of the channel for a few minutes, and scan the bed's surface with a laser sheet [Fig.~1(a), main document]. The intersection of the laser sheet with the bed appears in the view range of the camera used to track the traveling particles [Fig.~\ref{fig:laser_line}(a)]. The bed deforms this intersection in proportion to its own elevation $h$, allowing us to reconstruct the bed's cross-section, after identifying its position on the picture, $x_{laser}$ with the maximum of green light intensity [Fig.~\ref{fig:laser_line}(b)].

To calibrate this device, we first scan a channel filled with still milk, the whiteness of which generates a sharp laser line. The laser source and the camera are attached to a translating carriage, whose position is controlled with a step motor. We check that this 20-cm translation is horizontal, up to a slope of about $0.03\,$\% with respect to the milk's surface; we hereafter treat this value as the precision on the downstream slope measurement [Fig.~\ref{fig:laser_line}(c)]. Next, we add a fixed volume of milk into the channel, thus elevating the bath's surface by a known amount ($0.57\pm 0.08\,$cm), and scan its surface again. We repeat this procedure four times; the resulting shift of the laser line yields the conversion factor for the bed elevation:
\begin{equation}
  h \; [\mathrm{cm}] = ( 0.0031 \pm 0.0002 ) \, x_{laser} \; [\mathrm{pixel}] \, ,
\end{equation}
which corresponds to a laser inclination of $28.3^{\circ}$. The position of the laser line on the milk's surface varies across the channel by less than a pixel, and the resulting precision of the bed elevation measurement is below $30\,\mu$m---less than a grain diameter. The roughness of the bed therefore dominates the uncertainty on $h$. To average out some of this variability, we average our bed profiles over the 20$\,$cm-long excursion of the translating carriage.

\section{Stokes flow}

\begin{figure}
\centering
\includegraphics[width = \columnwidth]{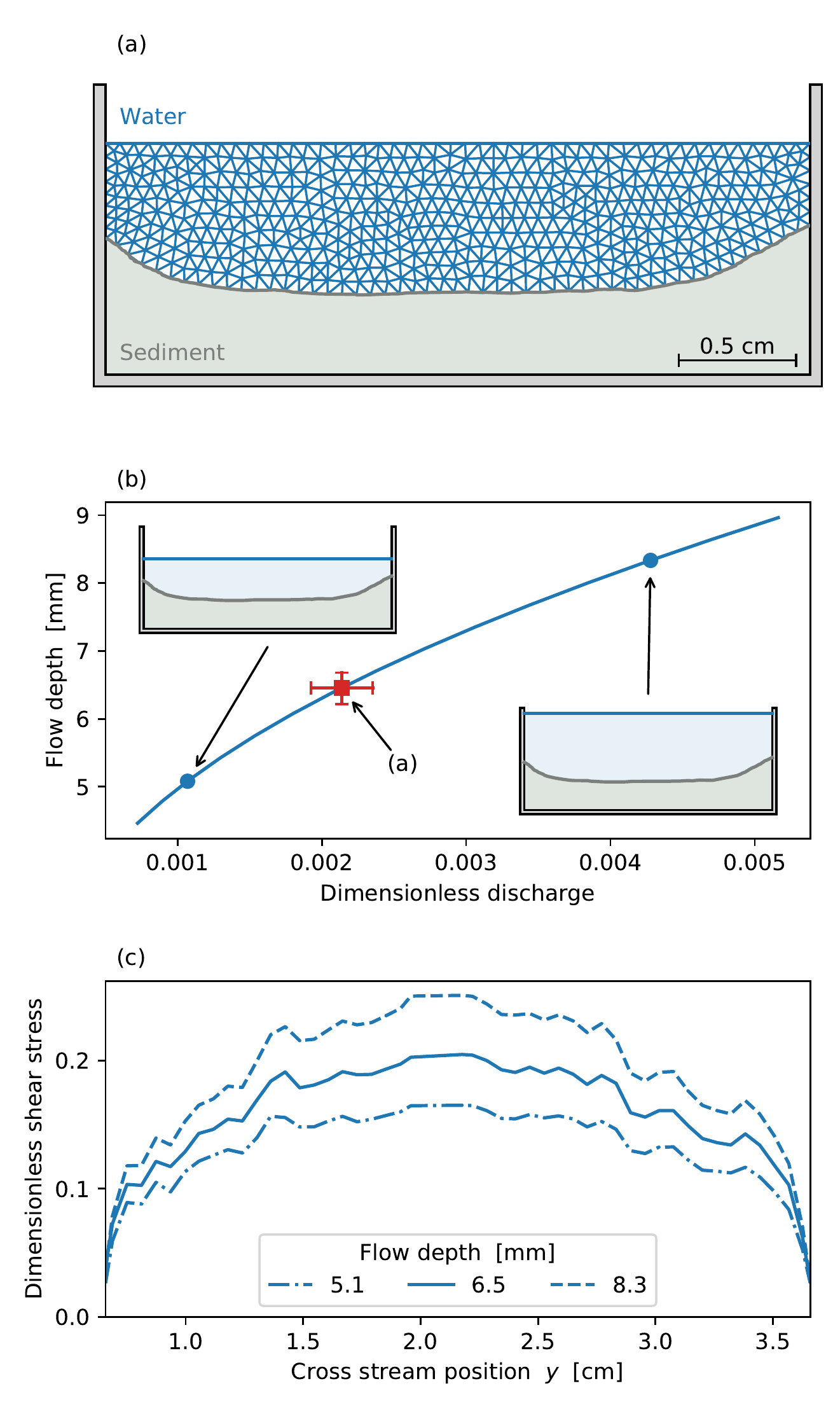}
\caption{Finite-elements simulation of the Stokes flow for run~\run. (a) Mesh used to calculate the flow field of Fig.~3(a) (main document). (b) Dependence of the flow depth on the dimensionless discharge. Solid line: numerical relation. Blue points: deepest and shallowest simulations in (c). Red square: best estimate for run~\run; error bars show uncertainty on dimensionless discharge, and resulting uncertainty on depth. (c) Influence of the maximum flow depth on the dimensionless shear stress.\label{fig:fluid_depth}}
\end{figure}

To produce the sediment transport law of Fig.~4(a) (main document), we need to estimate the shear stress induced by the fluid on the sediment bed. We do so by solving numerically the two-dimensional Stokes equation above the measured bed surface, using the software FreeFem++\footnote{\url{freefem.org}} \cite{MR3043640}.

In dimensionless form, the Stokes equation [equation (3), main document] reads
\begin{equation}
  \nabla^2 u_* = 1 \, ,
  \label{eq:dimless_Stokes}
\end{equation}
where $u_*$ is the downstream velocity field made dimensionless with $W^2 g S/\nu$, and space coordinates are made dimensionless with $W$. For a given flow depth and bed profile, solving Eq.~\eqref{eq:dimless_Stokes} numerically yields a velocity field and a dimensionless discharge $Q_{w*}$, with
\begin{equation}
  Q_{w*} = \dfrac{\nu Q_w}{g S W^4} = \iint u_* \, \mathrm{d}y \, \mathrm{d}z \, ,
\end{equation}
where integration is over the (dimensionless) cross section of the flow.

Unfortunately, we were not able to measure the elevation of the fluid surface with satisfactory precision using the laser scanner, mostly because of the fluctuation induced by traveling grains. Instead, we used the following procedure to estimate it. We first assume an elevation for the water surface, and generate a triangular-mesh approximation of the flow cross-section [Fig.~\ref{fig:fluid_depth}(a)]. We then solve Eq.~\eqref{eq:dimless_Stokes} using finite elements and compute the dimensionless discharge corresponding to this water elevation. Repeating these steps yields an approximate relation between discharge and flow depth [Fig.~\ref{fig:fluid_depth}(b)].

We use an electromagnetic flow meter to measure the fluid input into the channel, with a relative precision below 2$\,$\% (Kobold\footnote{\url{www.kobold.com}} Mik 0.5-10$\,$L$\,$min$^{-1}$). The fluctuations of the discharge during an experimental run are of the order of this precision. Considering the relative uncertainty on the slope $S$ and that on the viscosity $\nu$, we may estimate the dimensionless fluid discharge, $Q_{w*}$, with a relative precision of less than 10$\,$\% [red error bars on Fig.~\ref{fig:fluid_depth}(b)].
The numerical relation of Fig.~\ref{fig:fluid_depth}(b) then allows us to estimate the flow depth with an uncertainty of 0.2$\,$mm (less than 4$\,$\%). We may now evaluate how much the uncertainty on the flow depth affects the shear stress.

Turning our attention to Fig.~\ref{fig:fluid_depth}(c), we find that a 3-mm change of the flow depth affects the shear stress by up to 40$\,$\%; we therefore expect that the uncertainty on the computed shear stress, inherited from the uncertainty on the discharge, slope and viscosity, be less than 4$\,$\%---hardly a concern. Conversely, the actual variability of the bed elevation (about one grain diameter), would affect the shear stress by about 15$\,$\%. This estimate appears as an error bar in Fig.~3(b) (main document).

\bibliography{../main_document/Boltzmann_bedload_bibliography}

%% file: exemplary_experiment.tex
\newcommand{\run}{\#1}

%% file: Boltzmann_bedload.bbl
\begin{thebibliography}{37}%
\makeatletter
\providecommand \@ifxundefined [1]{%
 \@ifx{#1\undefined}
}%
\providecommand \@ifnum [1]{%
 \ifnum #1\expandafter \@firstoftwo
 \else \expandafter \@secondoftwo
 \fi
}%
\providecommand \@ifx [1]{%
 \ifx #1\expandafter \@firstoftwo
 \else \expandafter \@secondoftwo
 \fi
}%
\providecommand \natexlab [1]{#1}%
\providecommand \enquote  [1]{``#1''}%
\providecommand \bibnamefont  [1]{#1}%
\providecommand \bibfnamefont [1]{#1}%
\providecommand \citenamefont [1]{#1}%
\providecommand \href@noop [0]{\@secondoftwo}%
\providecommand \href [0]{\begingroup \@sanitize@url \@href}%
\providecommand \@href[1]{\@@startlink{#1}\@@href}%
\providecommand \@@href[1]{\endgroup#1\@@endlink}%
\providecommand \@sanitize@url [0]{\catcode `\\12\catcode `\$12\catcode
  `\&12\catcode `\#12\catcode `\^12\catcode `\_12\catcode `\%12\relax}%
\providecommand \@@startlink[1]{}%
\providecommand \@@endlink[0]{}%
\providecommand \url  [0]{\begingroup\@sanitize@url \@url }%
\providecommand \@url [1]{\endgroup\@href {#1}{\urlprefix }}%
\providecommand \urlprefix  [0]{URL }%
\providecommand \Eprint [0]{\href }%
\providecommand \doibase [0]{http://dx.doi.org/}%
\providecommand \selectlanguage [0]{\@gobble}%
\providecommand \bibinfo  [0]{\@secondoftwo}%
\providecommand \bibfield  [0]{\@secondoftwo}%
\providecommand \translation [1]{[#1]}%
\providecommand \BibitemOpen [0]{}%
\providecommand \bibitemStop [0]{}%
\providecommand \bibitemNoStop [0]{.\EOS\space}%
\providecommand \EOS [0]{\spacefactor3000\relax}%
\providecommand \BibitemShut  [1]{\csname bibitem#1\endcsname}%
\let\auto@bib@innerbib\@empty
%</preamble>
\bibitem [{\citenamefont {Bagnold}(1973)}]{bagnold1973nature}%
  \BibitemOpen
  \bibfield  {author} {\bibinfo {author} {\bibfnamefont {R.~A.}\ \bibnamefont
  {Bagnold}},\ }\href@noop {} {\bibfield  {journal} {\bibinfo  {journal} {Proc.
  R. Soc. Lond. A}\ }\textbf {\bibinfo {volume} {332}},\ \bibinfo {pages} {473}
  (\bibinfo {year} {1973})}\BibitemShut {NoStop}%
\bibitem [{\citenamefont {Mouilleron}\ \emph {et~al.}(2009)\citenamefont
  {Mouilleron}, \citenamefont {Charru},\ and\ \citenamefont
  {Eiff}}]{mouilleron2009inside}%
  \BibitemOpen
  \bibfield  {author} {\bibinfo {author} {\bibfnamefont {H.}~\bibnamefont
  {Mouilleron}}, \bibinfo {author} {\bibfnamefont {F.}~\bibnamefont {Charru}},
  \ and\ \bibinfo {author} {\bibfnamefont {O.}~\bibnamefont {Eiff}},\
  }\href@noop {} {\bibfield  {journal} {\bibinfo  {journal} {Journal of Fluid
  Mechanics}\ }\textbf {\bibinfo {volume} {628}},\ \bibinfo {pages} {229}
  (\bibinfo {year} {2009})}\BibitemShut {NoStop}%
\bibitem [{\citenamefont {Charru}\ \emph {et~al.}(2004)\citenamefont {Charru},
  \citenamefont {Mouilleron},\ and\ \citenamefont {Eiff}}]{charru2004erosion}%
  \BibitemOpen
  \bibfield  {author} {\bibinfo {author} {\bibfnamefont {F.}~\bibnamefont
  {Charru}}, \bibinfo {author} {\bibfnamefont {H.}~\bibnamefont {Mouilleron}},
  \ and\ \bibinfo {author} {\bibfnamefont {O.}~\bibnamefont {Eiff}},\
  }\href@noop {} {\bibfield  {journal} {\bibinfo  {journal} {Journal of Fluid
  Mechanics}\ }\textbf {\bibinfo {volume} {519}},\ \bibinfo {pages} {55}
  (\bibinfo {year} {2004})}\BibitemShut {NoStop}%
\bibitem [{\citenamefont {Lajeunesse}\ \emph {et~al.}(2010)\citenamefont
  {Lajeunesse}, \citenamefont {Malverti},\ and\ \citenamefont
  {Charru}}]{lajeunesse2010bed}%
  \BibitemOpen
  \bibfield  {author} {\bibinfo {author} {\bibfnamefont {E.}~\bibnamefont
  {Lajeunesse}}, \bibinfo {author} {\bibfnamefont {L.}~\bibnamefont
  {Malverti}}, \ and\ \bibinfo {author} {\bibfnamefont {F.}~\bibnamefont
  {Charru}},\ }\href@noop {} {\bibfield  {journal} {\bibinfo  {journal}
  {Journal of Geophysical Research: Earth Surface}\ }\textbf {\bibinfo {volume}
  {115}} (\bibinfo {year} {2010})}\BibitemShut {NoStop}%
\bibitem [{\citenamefont {Exner}(1925)}]{exner1925uber}%
  \BibitemOpen
  \bibfield  {author} {\bibinfo {author} {\bibfnamefont {F.~M.}\ \bibnamefont
  {Exner}},\ }\href@noop {} {\bibfield  {journal} {\bibinfo  {journal} {Akad.
  der Wiss in Wien, Math-Naturwissenschafliche Klasse, Sitzungsberichte, Abt
  IIa}\ }\textbf {\bibinfo {volume} {134}},\ \bibinfo {pages} {165} (\bibinfo
  {year} {1925})}\BibitemShut {NoStop}%
\bibitem [{\citenamefont {Coleman}\ and\ \citenamefont
  {Eling}(2000)}]{coleman2000sand}%
  \BibitemOpen
  \bibfield  {author} {\bibinfo {author} {\bibfnamefont {S.~E.}\ \bibnamefont
  {Coleman}}\ and\ \bibinfo {author} {\bibfnamefont {B.}~\bibnamefont
  {Eling}},\ }\href@noop {} {\bibfield  {journal} {\bibinfo  {journal} {Journal
  of Hydraulic Research}\ }\textbf {\bibinfo {volume} {38}},\ \bibinfo {pages}
  {331} (\bibinfo {year} {2000})}\BibitemShut {NoStop}%
\bibitem [{\citenamefont {Charru}\ and\ \citenamefont
  {Hinch}(2006)}]{charru2006ripple}%
  \BibitemOpen
  \bibfield  {author} {\bibinfo {author} {\bibfnamefont {F.}~\bibnamefont
  {Charru}}\ and\ \bibinfo {author} {\bibfnamefont {E.}~\bibnamefont {Hinch}},\
  }\href@noop {} {\bibfield  {journal} {\bibinfo  {journal} {Journal of Fluid
  Mechanics}\ }\textbf {\bibinfo {volume} {550}},\ \bibinfo {pages} {111}
  (\bibinfo {year} {2006})}\BibitemShut {NoStop}%
\bibitem [{\citenamefont {Devauchelle}\ \emph {et~al.}(2010)\citenamefont
  {Devauchelle}, \citenamefont {Malverti}, \citenamefont {Lajeunesse},
  \citenamefont {Josserand}, \citenamefont {Lagr{\'e}e},\ and\ \citenamefont
  {M{\'e}tivier}}]{devauchelle2010rhomboid}%
  \BibitemOpen
  \bibfield  {author} {\bibinfo {author} {\bibfnamefont {O.}~\bibnamefont
  {Devauchelle}}, \bibinfo {author} {\bibfnamefont {L.}~\bibnamefont
  {Malverti}}, \bibinfo {author} {\bibfnamefont {E.}~\bibnamefont
  {Lajeunesse}}, \bibinfo {author} {\bibfnamefont {C.}~\bibnamefont
  {Josserand}}, \bibinfo {author} {\bibfnamefont {P.-Y.}\ \bibnamefont
  {Lagr{\'e}e}}, \ and\ \bibinfo {author} {\bibfnamefont {F.}~\bibnamefont
  {M{\'e}tivier}},\ }\href@noop {} {\bibfield  {journal} {\bibinfo  {journal}
  {Journal of Geophysical Research: Earth Surface}\ }\textbf {\bibinfo {volume}
  {115}} (\bibinfo {year} {2010})}\BibitemShut {NoStop}%
\bibitem [{\citenamefont {Colombini}\ \emph {et~al.}(1987)\citenamefont
  {Colombini}, \citenamefont {Seminara},\ and\ \citenamefont
  {Tubino}}]{colombini1987finite}%
  \BibitemOpen
  \bibfield  {author} {\bibinfo {author} {\bibfnamefont {M.}~\bibnamefont
  {Colombini}}, \bibinfo {author} {\bibfnamefont {G.}~\bibnamefont {Seminara}},
  \ and\ \bibinfo {author} {\bibfnamefont {M.}~\bibnamefont {Tubino}},\
  }\href@noop {} {\bibfield  {journal} {\bibinfo  {journal} {Journal of Fluid
  Mechanics}\ }\textbf {\bibinfo {volume} {181}},\ \bibinfo {pages} {213}
  (\bibinfo {year} {1987})}\BibitemShut {NoStop}%
\bibitem [{\citenamefont {Ikeda}\ \emph {et~al.}(1981)\citenamefont {Ikeda},
  \citenamefont {Parker},\ and\ \citenamefont {Sawai}}]{ikeda1981bend}%
  \BibitemOpen
  \bibfield  {author} {\bibinfo {author} {\bibfnamefont {S.}~\bibnamefont
  {Ikeda}}, \bibinfo {author} {\bibfnamefont {G.}~\bibnamefont {Parker}}, \
  and\ \bibinfo {author} {\bibfnamefont {K.}~\bibnamefont {Sawai}},\
  }\href@noop {} {\bibfield  {journal} {\bibinfo  {journal} {Journal of Fluid
  Mechanics}\ }\textbf {\bibinfo {volume} {112}},\ \bibinfo {pages} {363}
  (\bibinfo {year} {1981})}\BibitemShut {NoStop}%
\bibitem [{\citenamefont {Johannesson}\ and\ \citenamefont
  {Parker}(1989)}]{johannesson1989linear}%
  \BibitemOpen
  \bibfield  {author} {\bibinfo {author} {\bibfnamefont {H.}~\bibnamefont
  {Johannesson}}\ and\ \bibinfo {author} {\bibfnamefont {G.}~\bibnamefont
  {Parker}},\ }\href@noop {} {\bibfield  {journal} {\bibinfo  {journal} {River
  meandering}\ }\textbf {\bibinfo {volume} {12}},\ \bibinfo {pages} {181}
  (\bibinfo {year} {1989})}\BibitemShut {NoStop}%
\bibitem [{\citenamefont {Liverpool}\ and\ \citenamefont
  {Edwards}(1995)}]{liverpool1995dynamics}%
  \BibitemOpen
  \bibfield  {author} {\bibinfo {author} {\bibfnamefont {T.}~\bibnamefont
  {Liverpool}}\ and\ \bibinfo {author} {\bibfnamefont {S.}~\bibnamefont
  {Edwards}},\ }\href@noop {} {\bibfield  {journal} {\bibinfo  {journal}
  {Physical review letters}\ }\textbf {\bibinfo {volume} {75}},\ \bibinfo
  {pages} {3016} (\bibinfo {year} {1995})}\BibitemShut {NoStop}%
\bibitem [{\citenamefont {Seminara}(2006)}]{seminara2006meanders}%
  \BibitemOpen
  \bibfield  {author} {\bibinfo {author} {\bibfnamefont {G.}~\bibnamefont
  {Seminara}},\ }\href@noop {} {\bibfield  {journal} {\bibinfo  {journal}
  {Journal of fluid mechanics}\ }\textbf {\bibinfo {volume} {554}},\ \bibinfo
  {pages} {271} (\bibinfo {year} {2006})}\BibitemShut {NoStop}%
\bibitem [{\citenamefont {Henderson}(1961)}]{henderson1961stability}%
  \BibitemOpen
  \bibfield  {author} {\bibinfo {author} {\bibfnamefont {F.~M.}\ \bibnamefont
  {Henderson}},\ }\href@noop {} {\bibfield  {journal} {\bibinfo  {journal}
  {Journal of the Hydraulics Division}\ }\textbf {\bibinfo {volume} {87}},\
  \bibinfo {pages} {109} (\bibinfo {year} {1961})}\BibitemShut {NoStop}%
\bibitem [{\citenamefont {Parker}\ \emph {et~al.}(2007)\citenamefont {Parker},
  \citenamefont {Wilcock}, \citenamefont {Paola}, \citenamefont {Dietrich},\
  and\ \citenamefont {Pitlick}}]{parker2007physical}%
  \BibitemOpen
  \bibfield  {author} {\bibinfo {author} {\bibfnamefont {G.}~\bibnamefont
  {Parker}}, \bibinfo {author} {\bibfnamefont {P.~R.}\ \bibnamefont {Wilcock}},
  \bibinfo {author} {\bibfnamefont {C.}~\bibnamefont {Paola}}, \bibinfo
  {author} {\bibfnamefont {W.~E.}\ \bibnamefont {Dietrich}}, \ and\ \bibinfo
  {author} {\bibfnamefont {J.}~\bibnamefont {Pitlick}},\ }\href@noop {}
  {\bibfield  {journal} {\bibinfo  {journal} {Journal of Geophysical Research:
  Earth Surface}\ }\textbf {\bibinfo {volume} {112}} (\bibinfo {year}
  {2007})}\BibitemShut {NoStop}%
\bibitem [{\citenamefont {Devauchelle}\ \emph {et~al.}(2011)\citenamefont
  {Devauchelle}, \citenamefont {Petroff}, \citenamefont {Lobkovsky},\ and\
  \citenamefont {Rothman}}]{devauchelle2011longitudinal}%
  \BibitemOpen
  \bibfield  {author} {\bibinfo {author} {\bibfnamefont {O.}~\bibnamefont
  {Devauchelle}}, \bibinfo {author} {\bibfnamefont {A.~P.}\ \bibnamefont
  {Petroff}}, \bibinfo {author} {\bibfnamefont {A.}~\bibnamefont {Lobkovsky}},
  \ and\ \bibinfo {author} {\bibfnamefont {D.~H.}\ \bibnamefont {Rothman}},\
  }\href@noop {} {\bibfield  {journal} {\bibinfo  {journal} {Journal of Fluid
  Mechanics}\ }\textbf {\bibinfo {volume} {667}},\ \bibinfo {pages} {38}
  (\bibinfo {year} {2011})}\BibitemShut {NoStop}%
\bibitem [{\citenamefont {Seizilles}\ \emph {et~al.}(2013)\citenamefont
  {Seizilles}, \citenamefont {Devauchelle}, \citenamefont {Lajeunesse},\ and\
  \citenamefont {M{\'e}tivier}}]{seizilles2013width}%
  \BibitemOpen
  \bibfield  {author} {\bibinfo {author} {\bibfnamefont {G.}~\bibnamefont
  {Seizilles}}, \bibinfo {author} {\bibfnamefont {O.}~\bibnamefont
  {Devauchelle}}, \bibinfo {author} {\bibfnamefont {E.}~\bibnamefont
  {Lajeunesse}}, \ and\ \bibinfo {author} {\bibfnamefont {F.}~\bibnamefont
  {M{\'e}tivier}},\ }\href@noop {} {\bibfield  {journal} {\bibinfo  {journal}
  {Physical Review E}\ }\textbf {\bibinfo {volume} {87}},\ \bibinfo {pages}
  {052204} (\bibinfo {year} {2013})}\BibitemShut {NoStop}%
\bibitem [{\citenamefont {Parker}(1978{\natexlab{a}})}]{parker1978self}%
  \BibitemOpen
  \bibfield  {author} {\bibinfo {author} {\bibfnamefont {G.}~\bibnamefont
  {Parker}},\ }\href@noop {} {\bibfield  {journal} {\bibinfo  {journal}
  {Journal of Fluid Mechanics}\ }\textbf {\bibinfo {volume} {89}},\ \bibinfo
  {pages} {109} (\bibinfo {year} {1978}{\natexlab{a}})}\BibitemShut {NoStop}%
\bibitem [{\citenamefont {Parker}(1978{\natexlab{b}})}]{parker1978self2}%
  \BibitemOpen
  \bibfield  {author} {\bibinfo {author} {\bibfnamefont {G.}~\bibnamefont
  {Parker}},\ }\href@noop {} {\bibfield  {journal} {\bibinfo  {journal}
  {Journal of Fluid mechanics}\ }\textbf {\bibinfo {volume} {89}},\ \bibinfo
  {pages} {127} (\bibinfo {year} {1978}{\natexlab{b}})}\BibitemShut {NoStop}%
\bibitem [{\citenamefont {Kovacs}\ and\ \citenamefont
  {Parker}(1994)}]{kovacs1994new}%
  \BibitemOpen
  \bibfield  {author} {\bibinfo {author} {\bibfnamefont {A.}~\bibnamefont
  {Kovacs}}\ and\ \bibinfo {author} {\bibfnamefont {G.}~\bibnamefont
  {Parker}},\ }\href@noop {} {\bibfield  {journal} {\bibinfo  {journal}
  {Journal of fluid Mechanics}\ }\textbf {\bibinfo {volume} {267}},\ \bibinfo
  {pages} {153} (\bibinfo {year} {1994})}\BibitemShut {NoStop}%
\bibitem [{\citenamefont {Chen}\ \emph {et~al.}(2009)\citenamefont {Chen},
  \citenamefont {Ma},\ and\ \citenamefont {Dey}}]{chen2009sediment}%
  \BibitemOpen
  \bibfield  {author} {\bibinfo {author} {\bibfnamefont {X.}~\bibnamefont
  {Chen}}, \bibinfo {author} {\bibfnamefont {J.}~\bibnamefont {Ma}}, \ and\
  \bibinfo {author} {\bibfnamefont {S.}~\bibnamefont {Dey}},\ }\href@noop {}
  {\bibfield  {journal} {\bibinfo  {journal} {Journal of Hydraulic
  Engineering}\ }\textbf {\bibinfo {volume} {136}},\ \bibinfo {pages} {311}
  (\bibinfo {year} {2009})}\BibitemShut {NoStop}%
\bibitem [{\citenamefont {Roseberry}\ \emph {et~al.}(2012)\citenamefont
  {Roseberry}, \citenamefont {Schmeeckle},\ and\ \citenamefont
  {Furbish}}]{roseberry2012probabilistic}%
  \BibitemOpen
  \bibfield  {author} {\bibinfo {author} {\bibfnamefont {J.~C.}\ \bibnamefont
  {Roseberry}}, \bibinfo {author} {\bibfnamefont {M.~W.}\ \bibnamefont
  {Schmeeckle}}, \ and\ \bibinfo {author} {\bibfnamefont {D.~J.}\ \bibnamefont
  {Furbish}},\ }\href@noop {} {\bibfield  {journal} {\bibinfo  {journal}
  {Journal of Geophysical Research: Earth Surface}\ }\textbf {\bibinfo {volume}
  {117}} (\bibinfo {year} {2012})}\BibitemShut {NoStop}%
\bibitem [{\citenamefont {Jenkins}\ and\ \citenamefont
  {Hanes}(1998)}]{jenkins1998collisional}%
  \BibitemOpen
  \bibfield  {author} {\bibinfo {author} {\bibfnamefont {J.~T.}\ \bibnamefont
  {Jenkins}}\ and\ \bibinfo {author} {\bibfnamefont {D.~M.}\ \bibnamefont
  {Hanes}},\ }\href@noop {} {\bibfield  {journal} {\bibinfo  {journal} {Journal
  of Fluid Mechanics}\ }\textbf {\bibinfo {volume} {370}},\ \bibinfo {pages}
  {29} (\bibinfo {year} {1998})}\BibitemShut {NoStop}%
\bibitem [{\citenamefont {Furbish}\ \emph {et~al.}(2012)\citenamefont
  {Furbish}, \citenamefont {Haff}, \citenamefont {Roseberry},\ and\
  \citenamefont {Schmeeckle}}]{furbish2012probabilistic}%
  \BibitemOpen
  \bibfield  {author} {\bibinfo {author} {\bibfnamefont {D.~J.}\ \bibnamefont
  {Furbish}}, \bibinfo {author} {\bibfnamefont {P.~K.}\ \bibnamefont {Haff}},
  \bibinfo {author} {\bibfnamefont {J.~C.}\ \bibnamefont {Roseberry}}, \ and\
  \bibinfo {author} {\bibfnamefont {M.~W.}\ \bibnamefont {Schmeeckle}},\
  }\href@noop {} {\bibfield  {journal} {\bibinfo  {journal} {Journal of
  Geophysical Research: Earth Surface}\ }\textbf {\bibinfo {volume} {117}}
  (\bibinfo {year} {2012})}\BibitemShut {NoStop}%
\bibitem [{\citenamefont {Lajeunesse}\ \emph {et~al.}(2017)\citenamefont
  {Lajeunesse}, \citenamefont {Devauchelle}, \citenamefont {Lachauss{\'e}e},\
  and\ \citenamefont {Claudin}}]{lajeunesse2017bedload}%
  \BibitemOpen
  \bibfield  {author} {\bibinfo {author} {\bibfnamefont {E.}~\bibnamefont
  {Lajeunesse}}, \bibinfo {author} {\bibfnamefont {O.}~\bibnamefont
  {Devauchelle}}, \bibinfo {author} {\bibfnamefont {F.}~\bibnamefont
  {Lachauss{\'e}e}}, \ and\ \bibinfo {author} {\bibfnamefont {P.}~\bibnamefont
  {Claudin}},\ }\href@noop {} {\bibfield  {journal} {\bibinfo  {journal}
  {Gravel-bed Rivers: Gravel Bed Rivers and Disasters, Wiley-Blackwell, Oxford,
  UK}\ ,\ \bibinfo {pages} {415}} (\bibinfo {year} {2017})}\BibitemShut
  {NoStop}%
\bibitem [{\citenamefont {Seizilles}\ \emph {et~al.}(2014)\citenamefont
  {Seizilles}, \citenamefont {Lajeunesse}, \citenamefont {Devauchelle},\ and\
  \citenamefont {Bak}}]{seizilles2014cross}%
  \BibitemOpen
  \bibfield  {author} {\bibinfo {author} {\bibfnamefont {G.}~\bibnamefont
  {Seizilles}}, \bibinfo {author} {\bibfnamefont {E.}~\bibnamefont
  {Lajeunesse}}, \bibinfo {author} {\bibfnamefont {O.}~\bibnamefont
  {Devauchelle}}, \ and\ \bibinfo {author} {\bibfnamefont {M.}~\bibnamefont
  {Bak}},\ }\href@noop {} {\bibfield  {journal} {\bibinfo  {journal} {Physics
  of Fluids}\ }\textbf {\bibinfo {volume} {26}},\ \bibinfo {pages} {013302}
  (\bibinfo {year} {2014})}\BibitemShut {NoStop}%
\bibitem [{\citenamefont {Nikora}\ \emph {et~al.}(2002)\citenamefont {Nikora},
  \citenamefont {Habersack}, \citenamefont {Huber},\ and\ \citenamefont
  {McEwan}}]{nikora2002bed}%
  \BibitemOpen
  \bibfield  {author} {\bibinfo {author} {\bibfnamefont {V.}~\bibnamefont
  {Nikora}}, \bibinfo {author} {\bibfnamefont {H.}~\bibnamefont {Habersack}},
  \bibinfo {author} {\bibfnamefont {T.}~\bibnamefont {Huber}}, \ and\ \bibinfo
  {author} {\bibfnamefont {I.}~\bibnamefont {McEwan}},\ }\href@noop {}
  {\bibfield  {journal} {\bibinfo  {journal} {Water Resources Research}\
  }\textbf {\bibinfo {volume} {38}},\ \bibinfo {pages} {17} (\bibinfo {year}
  {2002})}\BibitemShut {NoStop}%
\bibitem [{\citenamefont {Aussillous}\ \emph {et~al.}(2016)\citenamefont
  {Aussillous}, \citenamefont {Zou}, \citenamefont {Guazzelli}, \citenamefont
  {Yan},\ and\ \citenamefont {Wyart}}]{aussillous2016scale}%
  \BibitemOpen
  \bibfield  {author} {\bibinfo {author} {\bibfnamefont {P.}~\bibnamefont
  {Aussillous}}, \bibinfo {author} {\bibfnamefont {Z.}~\bibnamefont {Zou}},
  \bibinfo {author} {\bibfnamefont {{\'E}.}~\bibnamefont {Guazzelli}}, \bibinfo
  {author} {\bibfnamefont {L.}~\bibnamefont {Yan}}, \ and\ \bibinfo {author}
  {\bibfnamefont {M.}~\bibnamefont {Wyart}},\ }\href@noop {} {\bibfield
  {journal} {\bibinfo  {journal} {Proceedings of the National Academy of
  Sciences}\ }\textbf {\bibinfo {volume} {113}},\ \bibinfo {pages} {11788}
  (\bibinfo {year} {2016})}\BibitemShut {NoStop}%
\bibitem [{\citenamefont {Samson}\ \emph {et~al.}(1998)\citenamefont {Samson},
  \citenamefont {Ippolito}, \citenamefont {Batrouni},\ and\ \citenamefont
  {Lemaitre}}]{samson1998diffusive}%
  \BibitemOpen
  \bibfield  {author} {\bibinfo {author} {\bibfnamefont {L.}~\bibnamefont
  {Samson}}, \bibinfo {author} {\bibfnamefont {I.}~\bibnamefont {Ippolito}},
  \bibinfo {author} {\bibfnamefont {G.}~\bibnamefont {Batrouni}}, \ and\
  \bibinfo {author} {\bibfnamefont {J.}~\bibnamefont {Lemaitre}},\ }\href@noop
  {} {\bibfield  {journal} {\bibinfo  {journal} {The European Physical Journal
  B-Condensed Matter and Complex Systems}\ }\textbf {\bibinfo {volume} {3}},\
  \bibinfo {pages} {377} (\bibinfo {year} {1998})}\BibitemShut {NoStop}%
\bibitem [{\citenamefont {Munkres}(1957)}]{munkres1957algorithms}%
  \BibitemOpen
  \bibfield  {author} {\bibinfo {author} {\bibfnamefont {J.}~\bibnamefont
  {Munkres}},\ }\href@noop {} {\bibfield  {journal} {\bibinfo  {journal}
  {Journal of the society for industrial and applied mathematics}\ }\textbf
  {\bibinfo {volume} {5}},\ \bibinfo {pages} {32} (\bibinfo {year}
  {1957})}\BibitemShut {NoStop}%
\bibitem [{\citenamefont {Utter}\ and\ \citenamefont
  {Behringer}(2004)}]{utter2004self}%
  \BibitemOpen
  \bibfield  {author} {\bibinfo {author} {\bibfnamefont {B.}~\bibnamefont
  {Utter}}\ and\ \bibinfo {author} {\bibfnamefont {R.~P.}\ \bibnamefont
  {Behringer}},\ }\href@noop {} {\bibfield  {journal} {\bibinfo  {journal}
  {Physical Review E}\ }\textbf {\bibinfo {volume} {69}},\ \bibinfo {pages}
  {031308} (\bibinfo {year} {2004})}\BibitemShut {NoStop}%
\bibitem [{\citenamefont {Fenistein}\ \emph {et~al.}(2004)\citenamefont
  {Fenistein}, \citenamefont {van~de Meent},\ and\ \citenamefont {van
  Hecke}}]{fenistein2004universal}%
  \BibitemOpen
  \bibfield  {author} {\bibinfo {author} {\bibfnamefont {D.}~\bibnamefont
  {Fenistein}}, \bibinfo {author} {\bibfnamefont {J.~W.}\ \bibnamefont {van~de
  Meent}}, \ and\ \bibinfo {author} {\bibfnamefont {M.}~\bibnamefont {van
  Hecke}},\ }\href@noop {} {\bibfield  {journal} {\bibinfo  {journal} {Physical
  Review Letters}\ }\textbf {\bibinfo {volume} {92}},\ \bibinfo {pages}
  {094301} (\bibinfo {year} {2004})}\BibitemShut {NoStop}%
\bibitem [{\citenamefont {Debregeas}\ \emph {et~al.}(2001)\citenamefont
  {Debregeas}, \citenamefont {Tabuteau},\ and\ \citenamefont
  {Di~Meglio}}]{debregeas2001deformation}%
  \BibitemOpen
  \bibfield  {author} {\bibinfo {author} {\bibfnamefont {G.}~\bibnamefont
  {Debregeas}}, \bibinfo {author} {\bibfnamefont {H.}~\bibnamefont {Tabuteau}},
  \ and\ \bibinfo {author} {\bibfnamefont {J.-M.}\ \bibnamefont {Di~Meglio}},\
  }\href@noop {} {\bibfield  {journal} {\bibinfo  {journal} {Physical Review
  Letters}\ }\textbf {\bibinfo {volume} {87}},\ \bibinfo {pages} {178305}
  (\bibinfo {year} {2001})}\BibitemShut {NoStop}%
\bibitem [{\citenamefont {Hecht}(2012)}]{MR3043640}%
  \BibitemOpen
  \bibfield  {author} {\bibinfo {author} {\bibfnamefont {F.}~\bibnamefont
  {Hecht}},\ }\href@noop {} {\bibfield  {journal} {\bibinfo  {journal} {J.
  Numer. Math.}\ }\textbf {\bibinfo {volume} {20}},\ \bibinfo {pages} {251}
  (\bibinfo {year} {2012})}\BibitemShut {NoStop}%
\bibitem [{\citenamefont {Shields}(1936)}]{shields1936anwendung}%
  \BibitemOpen
  \bibfield  {author} {\bibinfo {author} {\bibfnamefont {A.}~\bibnamefont
  {Shields}},\ }\href@noop {} {\bibfield  {journal} {\bibinfo  {journal} {PhD
  Thesis Technical University Berlin}\ } (\bibinfo {year} {1936})}\BibitemShut
  {NoStop}%
\bibitem [{\citenamefont {Jop}\ \emph {et~al.}(2005)\citenamefont {Jop},
  \citenamefont {Forterre},\ and\ \citenamefont {Pouliquen}}]{jop2005crucial}%
  \BibitemOpen
  \bibfield  {author} {\bibinfo {author} {\bibfnamefont {P.}~\bibnamefont
  {Jop}}, \bibinfo {author} {\bibfnamefont {Y.}~\bibnamefont {Forterre}}, \
  and\ \bibinfo {author} {\bibfnamefont {O.}~\bibnamefont {Pouliquen}},\
  }\href@noop {} {\bibfield  {journal} {\bibinfo  {journal} {Journal of Fluid
  Mechanics}\ }\textbf {\bibinfo {volume} {541}},\ \bibinfo {pages} {167}
  (\bibinfo {year} {2005})}\BibitemShut {NoStop}%
\bibitem [{\citenamefont {Abramian}\ \emph {et~al.}(2018)\citenamefont
  {Abramian}, \citenamefont {Devauchelle},\ and\ \citenamefont
  {Lajeunesse}}]{abramian2018}%
  \BibitemOpen
  \bibfield  {author} {\bibinfo {author} {\bibfnamefont {A.}~\bibnamefont
  {Abramian}}, \bibinfo {author} {\bibfnamefont {O.}~\bibnamefont
  {Devauchelle}}, \ and\ \bibinfo {author} {\bibfnamefont {E.}~\bibnamefont
  {Lajeunesse}},\ }\href@noop {} {\bibfield  {journal} {\bibinfo  {journal}
  {Journal of Fluid Mechanics}\ } (\bibinfo {year} {2018})}\BibitemShut
  {NoStop}%
\end{thebibliography}%


\begin{thebibliography}{6}%
\makeatletter
\providecommand \@ifxundefined [1]{%
 \@ifx{#1\undefined}
}%
\providecommand \@ifnum [1]{%
 \ifnum #1\expandafter \@firstoftwo
 \else \expandafter \@secondoftwo
 \fi
}%
\providecommand \@ifx [1]{%
 \ifx #1\expandafter \@firstoftwo
 \else \expandafter \@secondoftwo
 \fi
}%
\providecommand \natexlab [1]{#1}%
\providecommand \enquote  [1]{``#1''}%
\providecommand \bibnamefont  [1]{#1}%
\providecommand \bibfnamefont [1]{#1}%
\providecommand \citenamefont [1]{#1}%
\providecommand \href@noop [0]{\@secondoftwo}%
\providecommand \href [0]{\begingroup \@sanitize@url \@href}%
\providecommand \@href[1]{\@@startlink{#1}\@@href}%
\providecommand \@@href[1]{\endgroup#1\@@endlink}%
\providecommand \@sanitize@url [0]{\catcode `\\12\catcode `\$12\catcode
  `\&12\catcode `\#12\catcode `\^12\catcode `\_12\catcode `\%12\relax}%
\providecommand \@@startlink[1]{}%
\providecommand \@@endlink[0]{}%
\providecommand \url  [0]{\begingroup\@sanitize@url \@url }%
\providecommand \@url [1]{\endgroup\@href {#1}{\urlprefix }}%
\providecommand \urlprefix  [0]{URL }%
\providecommand \Eprint [0]{\href }%
\providecommand \doibase [0]{http://dx.doi.org/}%
\providecommand \selectlanguage [0]{\@gobble}%
\providecommand \bibinfo  [0]{\@secondoftwo}%
\providecommand \bibfield  [0]{\@secondoftwo}%
\providecommand \translation [1]{[#1]}%
\providecommand \BibitemOpen [0]{}%
\providecommand \bibitemStop [0]{}%
\providecommand \bibitemNoStop [0]{.\EOS\space}%
\providecommand \EOS [0]{\spacefactor3000\relax}%
\providecommand \BibitemShut  [1]{\csname bibitem#1\endcsname}%
\let\auto@bib@innerbib\@empty
%</preamble>
\bibitem [{\citenamefont {Abramian}(2018)}]{theseAnais}%
  \BibitemOpen
  \bibfield  {author} {\bibinfo {author} {\bibfnamefont {A.}~\bibnamefont
  {Abramian}},\ }\emph {\bibinfo {title} {Self-organization of sediment
  transport in alluvial rivers}},\ \href@noop {} {Ph.D. thesis},\ \bibinfo
  {school} {{Universit{\'e} Sorbonne Paris Cit{\'e}}}, \bibinfo {address}
  {Institut de Physique du Globe de Paris} (\bibinfo {year} {2018}),\ \bibinfo
  {note} {partially in English}\BibitemShut {NoStop}%
\bibitem [{\citenamefont {Seizilles}\ \emph {et~al.}(2013)\citenamefont
  {Seizilles}, \citenamefont {Devauchelle}, \citenamefont {Lajeunesse},\ and\
  \citenamefont {M{\'e}tivier}}]{seizilles2013width}%
  \BibitemOpen
  \bibfield  {author} {\bibinfo {author} {\bibfnamefont {G.}~\bibnamefont
  {Seizilles}}, \bibinfo {author} {\bibfnamefont {O.}~\bibnamefont
  {Devauchelle}}, \bibinfo {author} {\bibfnamefont {E.}~\bibnamefont
  {Lajeunesse}}, \ and\ \bibinfo {author} {\bibfnamefont {F.}~\bibnamefont
  {M{\'e}tivier}},\ }\href@noop {} {\bibfield  {journal} {\bibinfo  {journal}
  {Physical Review E}\ }\textbf {\bibinfo {volume} {87}},\ \bibinfo {pages}
  {052204} (\bibinfo {year} {2013})}\BibitemShut {NoStop}%
\bibitem [{\citenamefont {Seizilles}\ \emph {et~al.}(2014)\citenamefont
  {Seizilles}, \citenamefont {Lajeunesse}, \citenamefont {Devauchelle},\ and\
  \citenamefont {Bak}}]{seizilles2014cross}%
  \BibitemOpen
  \bibfield  {author} {\bibinfo {author} {\bibfnamefont {G.}~\bibnamefont
  {Seizilles}}, \bibinfo {author} {\bibfnamefont {E.}~\bibnamefont
  {Lajeunesse}}, \bibinfo {author} {\bibfnamefont {O.}~\bibnamefont
  {Devauchelle}}, \ and\ \bibinfo {author} {\bibfnamefont {M.}~\bibnamefont
  {Bak}},\ }\href@noop {} {\bibfield  {journal} {\bibinfo  {journal} {Physics
  of Fluids}\ }\textbf {\bibinfo {volume} {26}},\ \bibinfo {pages} {013302}
  (\bibinfo {year} {2014})}\BibitemShut {NoStop}%
\bibitem [{\citenamefont {Munkres}(1957)}]{munkres1957algorithms}%
  \BibitemOpen
  \bibfield  {author} {\bibinfo {author} {\bibfnamefont {J.}~\bibnamefont
  {Munkres}},\ }\href@noop {} {\bibfield  {journal} {\bibinfo  {journal}
  {Journal of the society for industrial and applied mathematics}\ }\textbf
  {\bibinfo {volume} {5}},\ \bibinfo {pages} {32} (\bibinfo {year}
  {1957})}\BibitemShut {NoStop}%
\bibitem [{\citenamefont {Ancey}\ and\ \citenamefont
  {Heyman}(2014)}]{ancey2014microstructural}%
  \BibitemOpen
  \bibfield  {author} {\bibinfo {author} {\bibfnamefont {C.}~\bibnamefont
  {Ancey}}\ and\ \bibinfo {author} {\bibfnamefont {J.}~\bibnamefont {Heyman}},\
  }\href@noop {} {\bibfield  {journal} {\bibinfo  {journal} {Journal of Fluid
  Mechanics}\ }\textbf {\bibinfo {volume} {744}},\ \bibinfo {pages} {129}
  (\bibinfo {year} {2014})}\BibitemShut {NoStop}%
\bibitem [{\citenamefont {Hecht}(2012)}]{MR3043640}%
  \BibitemOpen
  \bibfield  {author} {\bibinfo {author} {\bibfnamefont {F.}~\bibnamefont
  {Hecht}},\ }\href@noop {} {\bibfield  {journal} {\bibinfo  {journal} {J.
  Numer. Math.}\ }\textbf {\bibinfo {volume} {20}},\ \bibinfo {pages} {251}
  (\bibinfo {year} {2012})}\BibitemShut {NoStop}%
\end{thebibliography}%
